\documentclass{article}
\usepackage[margin=1in]{geometry}

\usepackage{graphicx} 
\usepackage{pdfpages}
\usepackage{tikz}
\usepackage{pgf-pie}
\usepackage{booktabs}
\usepackage{longtable}
\usepackage{tabularray}
\usepackage{comment}
\usepackage{enumitem}
\usepackage{makecell}
\usepackage{rotating}

\usepackage{url}
\usepackage{todonotes}

\usepackage{setspace}

\setlist[itemize]{noitemsep}

\usepackage[linktocpage=true]{hyperref}
\hypersetup{
    colorlinks=true,
    linkcolor=blue,
    filecolor=magenta,      
    urlcolor=cyan,
    pdftitle={Overleaf Example},
    pdfpagemode=FullScreen,
}

\title{NSF CSSI-CyberTraining-SCIPE PI Meeting }

\begin{document}

\newpage
\pagenumbering{gobble}

\newpage


\vspace*{\fill}
\begin{center}
\Large
2024 NSF CSSI-Cybertraining-SCIPE PI Meeting \\
August 12 to 13, 2024, Charlotte, NC \\
\vspace{6pt}
\small{Link to meeting:  \url{https://confmeet.github.io/ccs2024/}}
\end{center}

\begin{center}
\textit{Co-Chairs:} \\
Abani Patra (Tufts University), Mary Thomas (University of California San Diego) 
\vspace{6pt}

\textit{Program Committee Members: }\\
Elias Bou-Harb (Louisiana State University, CSSI), 
Jeffrey Carver (University of Alabama Tuscaloosa, CSSI), 
Yuebin Guo (Rutgers University, CSSI), 
Ratnesh Kumar (Iowa State University, CSSI), 
Julien Langou (University of Colorado Denver, CSSI), 
Guoyu Lu (SUNY Binghamton, CSSI), 
Vivak Patel (University of Wisconsin, CSSI), 
Marianna Safronova (University of Delaware, CSSI), 
Isla Simpson (NSF NCAR, CSSI), 
Dhruva Chakravorty (Texas A\&M, CT), 
Jane Combs (University of Cincinnati, CT), 
Hantao Cui (Oklahoma State, CT), 
Sushil Prasad (UT San Antonio, CT), 
Adnan Rajib (UT Arlington, CT), 
Susan Rathbun (San Diego Super Computer Center, CT), 
Erik Saule (University of North Carolina -- Charlotte, CT), 
Isla Simpson (UCAR, CT), 
Alan Sussman (University of Maryland, CT), 
Shaowen Wang (University of Illinois Urbana-Champaign, CT), 
Sarina (Zhe) Zhang (Texas A\&M, CT)
\vspace{6pt}

\textit{Additional Contributors to Sessions and Report Preparation: }\\
Katie Antypas (NSF), 
Ritu Arora (Wayne State), 
Ben Brown (DOE/OASCR), 
Varun Chandola (NSF), 
Daniel Crawford (Virgina Tech), 
Ian Foster (U. Chicago/ Argonne), 
Dave Hart (NCAR), 
Mike Heroux (Sandia), 
Mary Ann Leung (Sustainable Horizons Institute), 
Benjamin Lynch (U Minn),
Dan Negrut (UW Madison), 
D. K. Panda (OSU), 
Manish Parashar (Utah),
Melissa Kline Struhl (MIT),
George K. Thiruvathukal (Loyola University Chicago)

\vspace{6pt}
\end{center}

\textit{Abstract} The second annual NSF/OAC CSSI/CyberTraining and related programs PI meeting was held August 12–13 in Charlotte, NC, with participation from PIs or representatives of all major awards. Keynotes, panels, breakouts, and poster sessions allowed PIs to engage with each other, NSF staff, and invited experts. The 286 attendees represented 292 awards across CSSI, CyberTraining, OAC Core, CIP, SCIPE CDS\&E, and related programs, and presented over 250 posters. This report documents the meeting’s structure, findings, and recommendations, offering a snapshot of current community perspectives on cyberinfrastructure (CI). A key takeaway is a vibrant, engaged community advancing science through CI. AI-driven research modalities complement established HPC and data-centric tools. Workforce development efforts align well with the CSSI community.



\paragraph{Disclaimer.}
This report is based on activities supported by the National Science Foundation under award numbers 2435580 and 2434556.
Any opinions, findings, conclusions or recommendations expressed herein are those of the authors and do not necessarily reflect the views of the National Science Foundation.

\paragraph{Acknowledgements} We acknowledge the efforts of the following without whose contributions above and beyond the call of duty this meeting would simply not have happened!
\begin{description}
    \item[{\cal{o}}] Our NSF program directors who supported us with their time and resources.
    \item[{\cal{o}}] Susan Rathbun, Jordan Wilkinson and Meghan Rodriguez whose efforts were crucial to the organization and smooth running of the meeting.
\end{description}


\vspace*{\fill}
\newpage

\begin{spacing}{0.9}\tableofcontents\end{spacing}
\newpage




\newpage
\pagenumbering{arabic}
\newpage


\section{Executive Summary}
\label{sec:exec-sum}
\subsection{Summary}
The NSF/OAC CSSI/Cybertraining and related programs PI meeting was successfully conducted August 12-13 in Charlotte, NC\cite{CyTr-CSSI-PIMtg24}. The meeting was attended by PIs and/or representatives of all major awards from both of these programs. This was the second joint CSSI and CyberTraining meeting \cite{CyTrPIMtg23}. Keynote, panels, breakouts and required posters provided PIs opportunites to learn from each other, NSF personnel and distinguished invitees. The purpose of this report is to document meeting processes and record findings and recommendations that provide a snapshot of the community thinking about these programs and cyberinfrastructure current state and futures.

The primary finding from the different activities at the meeting and inputs provided is a thriving and engaged community of scientists  at the intersection of  cyberinfrastructure(CI) and science building, sustaining and  CI for science. Newer research modalities driven by AI complement the robust HPC and data driven CI enabled scientific tools. Training and workforce development communities integrate well with the CSSI community.



\subsection{Future Directions  }

\paragraph{Overall Future Directions}
Attending PIs used part of the breakout times to define  desired future directions. In a desired future there will be a skilled and well trained workforce capable of enabling and sustaining the ML/AI transformation, supported by low barrier access to tools, data, flexible and scalable storage solutions, science workflows and high-performance computing resources through advanced CI. This will enable several high priority scientific use cases in fields requiring large-scale data processing and long-term storage like environmental modeling, genomic research, and AI-driven simulations in physics (secs. \ref{sec:sec:breakout_1-fut}, \ref{sec:breakout_2-fut} and \ref{sec:breakout_4-fut} for details).
Beyond these domains the use of AI in science is still in its infancy and good CI can enable use of AI in many fields (sec. \ref{sec:breakout_4-fut}). 
Effective translation of science advances to tools requires a complete ecosystem(hardware, tools and people) and culture enabling attributes like reproducibility in science done with computing (sec. \ref{sec:breakout2_2-fut} ). Successful CI will therefore have incentives for developers supported write better scientific software across domains (sec. \ref{sec:breakout2_4-fut}).
While, the programs in this cohort are developing distinct communities that focused on  critical albeit different aspects of computationally enabled research integration and strengthening structures within and across communities can enable better science (secs. \ref{sec:breakout2_5-fut} and \ref{sec:breakout2_6-fut}).
Finally, a truly desired outcome is the recognition and formalization of the roles of Research Software engineers as co-equal to research staff, closely integrated and integral to making the science happen
- through both career recognition, citations, as well as stable career paths (sec. \ref{sec:breakout_3-fut}).














\subsection{Recommendations  }

\paragraph{Overall Recommendations}
\begin{enumerate}  
   \item[{\bf R1}] It is recommended that the organizing committee is appointed early and  meeting planning  and communication to attendees start 6-9 months before meeting time. 
    Cost (registration and travel) and convenience of attending (travel, hotel and dates) are high priorities for attendees.
   Resources from small registration fee were crucial to hosting meeting seamlessly with many  unanticipated costs or costs that are difficult to charge to NSF grants.
   \item[{\bf R2}]Poster sessions are well received and provide meaningful opportunities for networking and sharing of best practices among diverse groups of researchers from many areas and backgrounds and NSF personnel. Therefore, the experience at posters should be high priority.  
    Breakouts are effective means of generating community input with good feedback on diverse topics   but good breakouts need structure and time.
 Panels are less effective at holding PI's attention though considerable expenditure is entailed in getting good panelists.
 \item[{\bf R3}] Investments leading to low barrier access to high end CI, innovative approaches to provisioning training on ML/AI methods, supporting standardization and interoperability and long term engagement of necessary personnel are needed for enabling the ML/AI transformation of science. Furthermore, there needs to be clear domain-specific standards for data collection, data curation and data
sharing such that the resulting data can be used to train AI including foundation models for use integral to advanced  CI. Significant investment is required to set up and operate such data frameworks.

 \item[{\bf R4}] Small EAGER like grants for hardware maintenance, expanded training in supporting heterogeneous computing resources, funding for improving AI hardware effectiveness and encouragement for local, regional and national    ``condo" models for  computing hardware more accessible for the sciences.  
 \item[{\bf R5}] The critical role of RSEs in the CI ecosystem needs to be carefully supported through diverse funding models, treating RSE support as integral to major CI investments. Matching of RSEs to domain sciences builds expertise and beneficial relationships.
 \item[{\bf R6}] Integrating CI specialists and domain science experts in training opportunities can break silos. Including graduate students and early career scientists in such training is desirable.
 \item[{\bf R7}] To support translation of science advances to tools, encourage mechanisms for providing credit for software and tools like DOIs for software. Encourage other directorates to value science tools even if supported by OAC division. Provide sustainability grants for science and emphasize a culture of reproducibility. It is also recommended that a clearer path for integrating CSSI software into ACCESS and future NSF supported resources be created.
 \item[{\bf R8}] Community building requires refined and expanded metrics developed in collaboration with social scientists on both quantitative and qualitative aspects of community engagement for better measurement of community sentiment and long term engagement.
 \item[{\bf R9}] Advanced CI needs community-driven and innovative curricula to   adapt
to rapidly changing technology with multiple delivery formats from reskilling like the Software Carpentries to diverse targets ranging from K-12, community colleges and current CI
professional from academia and industry. 

 \item[{\bf R10}]
Encourage inter-directorate and industry Collaboration    with other NSF directorates, such as EDU and domain science  directorates. In particular partnering with the
NSF TIP directorate and industry partners to develop tools and training which is valuable to both science professionals and industry
for staff upskilling.
\end{enumerate}


\paragraph{Detailed Recommendations}
\begin{itemize}  
    \item {\bf Meeting Organization } and Running
From ``Notes for Next year’s (2025) planning" document compiled as we organized and ran the meeting.
\begin{itemize}
    \item Start planning earlier! This year's meeting decision making was dominated by the deadlines.
    \item List the NSF program affiliation on the name badge (CSSI, SCIPE, Cybertraining, etc)
    \item Post web info in more places (add to slides)
    \item Bell or some way to get people to leave poster session and go to next sessions
    \item Poster session was very loud (good) - maybe leave more space between poster rows so conversation is easier.
    \item CaRCC professionalization and workforce development presentation?
    \item Short URLs for breakout sessions scribes 
    \item Full A/V in breakout rooms
    \item Extend invites to the team members beyond PIs- people actually doing much of the work - who have few opportunities to engage. 
    \item Chair needs to be able to coordinate the event and have professional event management and conferencing people to execute the event (they can be funded via the conference proposal).
    \item Recommend that the event is held in the city of the chair(s)
    \item Find way to get PI + CoPIs for outreach/mailing list
    \item Setup long-term DN?
\end{itemize}
\item {\bf Enabling the ML/AI Transformation of Science Discovery and Innovation}(see sec: \ref{sec:breakout_1-rec})
\begin{itemize}
    \item Investments leading to low barrier access to high end CI
    \item Support innovative approaches to training on ML/AI methodologies, workflow development, and infrastructure usage.  
    \item Investments in standardized data formats and robust frameworks to enhance data quality and interoperability
    \item Develop programs that address the need for long-term support for personnel
    \item Ensure that AI resources are inclusive and accessible to a broad range of researchers
\end{itemize}
\item {\bf Access to hardware resources}(see sec: \ref{sec:breakout_2-rec} for details)
\begin{itemize}
    \item Introduce Small, EAGER-like Awards for Hardware Maintenance and Upgrades
    \item Enhance Support for Scalable Data Storage Solutions
    \item Expand Training Programs for Heterogeneous Computing and Interoperability
    \item Develop Campus-Level ``Condo Model” for Shared Hardware Resources
    \item Fund Studies on Improving AI Hardware Energy Efficiency
    \item Establish a Long-Term Hardware Support Fund
    \item Develop a Sustainable Data Management Strategy
    \item Establish Permanent NSF-Funded Training Centers
    \item Promote Regional or National ``Condo Models” for Shared Resources
    \item Prioritize Energy Efficiency in Hardware Funding Decisions: Make energy efficiency a key
    
\end{itemize}
\item{\bf Role of RSEs and Support Staff} (see \ref{sec:breakout_3-rec} for details).
\begin{itemize}
    \item Diverse funding models for stable support 
    \item Invest in RSEs integral to major CI investments  
    \item Match RSE expertise and experience to domain science
    \item Support teams of RSEs with different levels and domains of expertise  
\end{itemize}

\item {\bf Designing Specific AI Tools for Science Discovery and Innovation} (see \ref{sec:breakout_4-rec} for details)
\begin{itemize}
    \item Early career AI training opportunities  
    \item Support for Shared data and sharing mechanisms  
    \item Infrastructure support for domain specific AI tools  
\end{itemize}

\item {\bf Training Resources} (see \ref{sec:breakout_5-rec} for details)
\begin{itemize}
    \item Train research computer scientists to work with domain experts (mainly grad students). Domain experts know how to compute pre-HPC (e.g. Office applications on a laptop), and need help to use HPC with AI. 
    \item Train research software engineers to be experts in computer science and familiar with the target domains.
    \item Develop a workforce that spans undergraduate students, who are not experts in the domain-science, but, who are nimble at learning software tools and can provide maintenance for installing needed packages
    \item Teach community to overcome silos in goal setting (performance vs actual research goals.) and set realistic objectives
    \item Train users to identify the limitations of ``black-box" AI: what worked, why, etc
    \item Delivering of training - look to non-traditional ways that training is deployed
    \item Provide a searchable mechanism within domain-specific areas so others can learn more quickly and identify good tools to use 
    \item Organize frequent workshops targeting PIs, domain scientists, RSEs, and computer scientists to learn novel methods and keep up with evolving AI trends
\end{itemize}

\item {\bf Translating Science Advances into CSSI Tools: from Papers to Software } (see \ref{sec:breakout2_2-rec} for details)
\begin{itemize}
    \item {\bf Encourage mechanisms for providing credit for software and tools.} Require DOI’s for publicly available software/data resources; citations and papers in appropriate journals. Encourage inclusion of tools and their usage in grant reporting across both CSSI and domain science grants to promote recognition of software development activity.
    \item {\bf Sustainability grants for software tools for science.} The need grants for sustainability to keep software working was reinforced. Recent addition of the sustainability track was recognized. Such support ``keeps the lights on and provides a base for rapid adoption and sustained use of computing driven science.
    \item {\bf Emphasize science culture with reproducibility of science advances and reuse of available scientific tools.} Reproducibility initiatives that promote widespread sharing of scientific tools based on science advances help.
    \item {\bf Balanced funding sources.} Certain funding sources should prioritize students while others support research scientists and engineers who can develop and maintain professional quality software.
    \item Bridging the gap between community stakeholders and tool developers is crucial for interdisciplinary collaboration. There is often a lack of recognition for these efforts and the challenges of balancing specialized tools with broader generalization. Tools may not be immediately appreciated by the community, and reliance on student teams poses risks if they are not sustained beyond graduation. Funding priorities sometimes favor students over research scientists and engineers, impacting the development and maintenance of high-quality software. Long-term support, potentially modeled after National Labs, is important for scaling and updating software. Globus is highlighted as an example of effective software support through a freemium model that balances open access with premium features.
\end{itemize}

\item {\bf Integrating AI/Foundation Models into CSSI} (see \ref{sec:breakout2_3-rec} for details)
\begin{itemize}
\item To address the challenges of using foundation models for cyberinfrastructure for sustained scientific innovation, there are two primary recommendations.
\item there needs to be clear domain-specific standards for data collection, data curation and data sharing. The resulting data can be used to train foundation models for use in CI. 
\item Second, there needs to be a substantial investment in developing this data infrastructure, and for training individuals to maintain, contribute to, and use this data infrastructure.
\end{itemize}

\item {\bf Integrating New Hardware into CSSI Software} (see \ref{sec:breakout2_4-rec} for details)
\begin{itemize}
\item Change the perception of POSE / CSSI-sustainability grants from “end of funding” to “next step in maintaining a healthy software base”.   
\item Create systems in reporting to value software like papers, track the requested metrics around adoption, etc., and hold PIs *accountable* for them (e.g. impacts on renewal or future SW funding). 
\item Spread the OAC culture of valuing software across NSF. 
\item Create a clearer path to transition CSSI-created software products throughout the NSF CI ecosystem (ACCESS, LCCF, etc.). 
\item Can we work with institutions to create an ecosystem to help support research software? (e.g. underlying expectations in release procedures, network infrastructure, etc.).  
\end{itemize}

\item {\bf Community Building \& Measurable Broader Impacts} (see \ref{sec:breakout2_5-rec} for details)
\begin{itemize}
\item Refine and Expand Metrics: Collaborate with social scientists to develop new metrics that capture both quantitative and qualitative aspects of community engagement. Consider what tools can analyze community sentiment, diversity, and long-term engagement trends. Allow communities to define what matters to them.
\item Innovate Broader Impact Strategies: Encourage original thinking in broader impact planning. Ensure appropriate budget allocations for long-term community engagement and sustainability initiatives. Explore the use of unstructured surveys to gain deeper insights.
\item Strengthen Privacy and Legal Compliance: Work with legal experts to ensure all tools and methods used in community-building are compliant with Federal, State and privacy laws. Consider adopting practices like partial name storage or anonymization to protect personally identifiable information while maintaining functionality.  Community leads must make a habit of requesting consent for all personal information collected and leveraging their institutional review boards (IRBs).
\item Invest in Expertise: There is a need for a coordinated effort that brings expertise together in a cohesive manner. Mechanisms could include hiring or consulting with social scientists or other specialists in community engagement to guide the development of community-building initiatives. Provide training and resources to investigators to help them better understand and implement these strategies.
\end{itemize}

\item {\bf Sustainability \& Continuing Training} (see \ref{sec:breakout2_6-rec} for details)
\begin{itemize} 
\item Structured Organization and Community Building:  Develop an 'Alliance' model which fosters broader partnerships, bringing projects together under collaborative umbrellas, similar to Research Coordination Networks (RCNs), BigData Hubs, AI institutes. Alliances and hubs, which may be discipline or regionally focused, will share resources and best practices via a central repository to avoid duplication of efforts.  Regional collaboration can foster the sharing of cybertraining-proficient people for scalability. Create a digital library of repository materials and share formal curricula through this federated repository.
\item Community-driven and Innovative Curriculum:  The cyberinfrastructure community must adapt to rapidly changing technology by ensuring that materials and delivery methods remain up to date and relevant.  Community-driven curriculum and topics are important to maintain our competitive advantage.  Programs similar to the Software Carpentries program should be considered to teach cyberinfrastructure to scientists, specifically cloud based infrastructure as a service.  Programs should expand their target audiences to include K-1, community colleges and current CI professional from academia and industry.  Programs should be considered which teach professional skills in addition to technical skills, include effective communication, project management and cybersecurity/privacy awareness.
\item Inter-directorate and Industry Collaboration:  It was suggested that the development of two-way relationships with other NSF directorates, such as EDU and domain directorates, would be useful, specifically incorporating EDU pedagogy research into CyberTraining awards.  A strong recommendation from this year's participants and last year's participants is to partner with the NSF TIP directorate and industry partners to develop training which is valuable to industry for staff upskilling providing an opportunity for subsidies and sustainability in addition to other opportunities.
\end{itemize}

\end{itemize}  

\subsection{Broader Impacts  }
The PI meeting brought together leading experts in the CSSI and broader communities to discuss
and share innovations and best practices of developing and sustaining cyberinfrastructure over
time. These new and reinforced collaborations will build the capacity for sustainable cyberinfrastructure services that
can enhance productivity and accelerate innovation in science and engineering and will
significantly contribute to increasing the impact of the output of NSF and specifically OAC’s programs. The sharing of best practices around focused panel discussions and posters will allow many impacts outside the primary disciplines for which the awards had been made.

Careful consideration was made in selecting speakers and panelists to ensure they came from
a diverse set of backgrounds     consistent with community best practices.
\label{sec:broader-impacts}

\section{Meeting Overview, Goals, Planning and Execution} 
\label{sec:meeting-overview}
\subsection{Meeting Goals {\it  }}
 The CSSI PI meeting  supports the community building efforts of past Software
Infrastructure for Sustained Innovation (SI2) and CSSI workshops. CSSI PI meetings provide a
forum for PIs to share technical information about their projects with each other, NSF program
directors and others; to explore innovative topics emerging in the software and data infrastructure
communities; to discuss and learn about best practices across projects; and to stimulate new
ideas of achieving software and data sustainability and ensuring a diverse pipeline of CI
researchers and professionals. PIs also provide valuable feedback to the program on emerging
opportunities and challenges. For this year in addition to the orginal goals of building and sustaining good cyberinfrastructure we also expected to see the effect of the AI driven transformation on the NSF CSSI/Cybertraining research community. The PI meetings have resulted in the formation of many new collaborations along with a sharing of best practices.

\subsection{Committee and Meeting Organization  }

\subsection{Meeting Content {\it  }}

\begin{figure}
\includegraphics[width=6.5in]{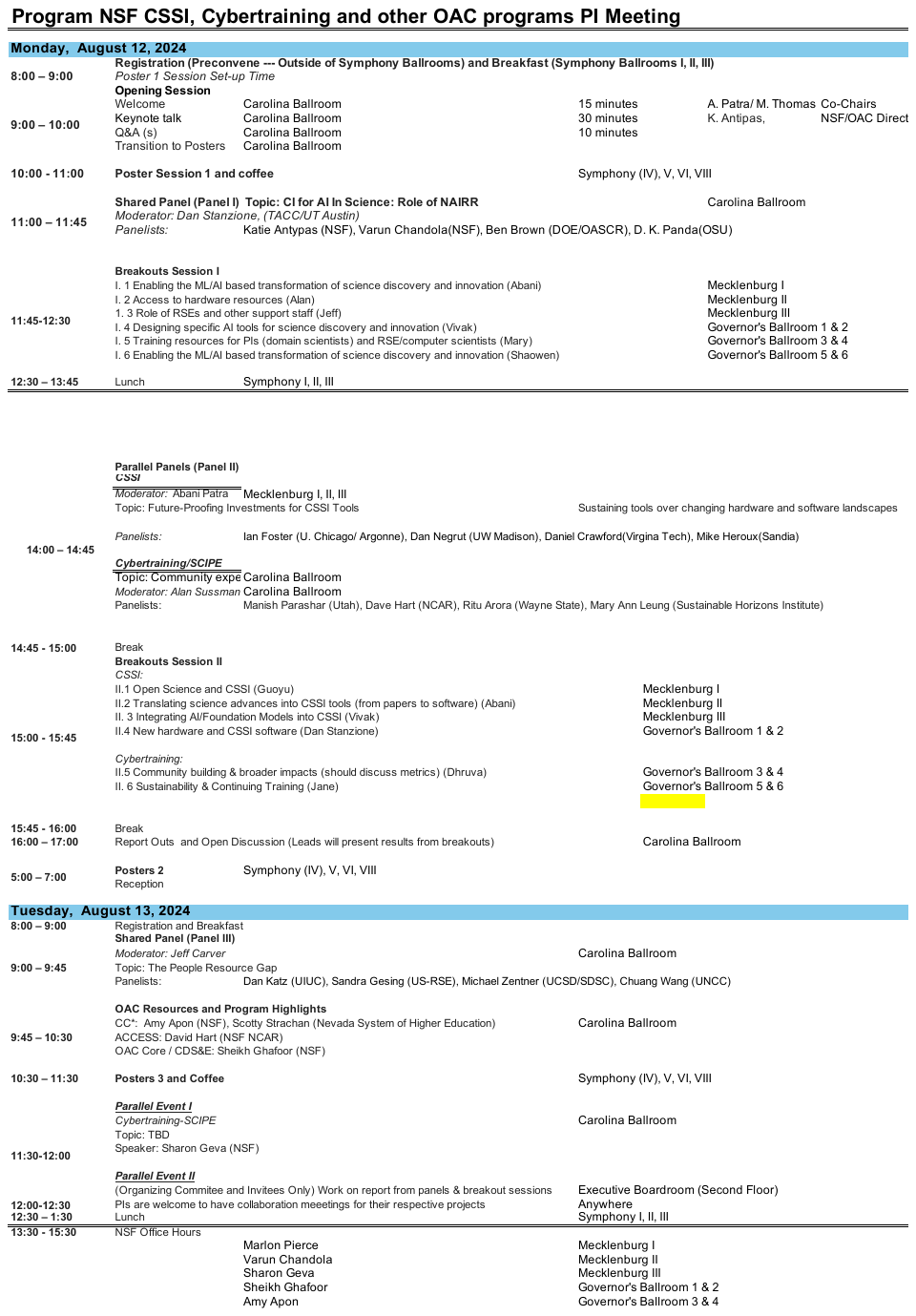}
\label{fig:program}
\caption{Program and Schedule}
\end{figure}

The final program provided a strong overview of CSSI, Cybertraining, SCIPE and had representation from other NSF/OAC programs like ACCESS and CC* (see program in Figure \ref{fig:program}). A keynote, two sets of shared panels (for all participants) and one set of parallel panels (one for CSSI and one for Cybertraining) were complemented with two sets of breakouts and three sets of poster sessions(required for every funded project).

A keynote talk by the director of NSF/OAC anchored the program and provided a comprehensive overview of NSF/OAC programs and plans \footnote{Travel disruptions moved this talk to a later time to no detriment in the overall flow of the meeting}. The panels provided active discussion from a distinguished set of invited community leaders (see details in subsequent sections). The breakouts provided an opportunity for every participant to provide input that we have processed into findings and recommendations that are very insightful not only into successes and pain points of existing programs but also provide some guidelines for future interests of participants.
The posters (open access in figshare; see appendix for list of posters) are at the heart of this meeting. The level of participation was extremely high and the poster rooms were busy and hosted many productive conversations supporting networking and future collaborations. 

\section{Community Feedback  }
\label{sec:comm-feedback}
We conducted an anonymous post-meeting survey of all attendees to get feedback about different aspects of the meeting, including pre-meeting communication.
We received 163 responses from the 284 registrants (some of whom were unable to attend the meeting), which is a response rate of at least $57\%$.
We had 116 responses from those affiliated with the CSSI program; 44 responses from those affiliated with the Cybertraining program; 10 responses from those affiliated with the SCIPE program; and the remaining responses came from individuals affiliated with other programs.\footnote{These values include overlaps as individuals can be affiliated with more than one program.}

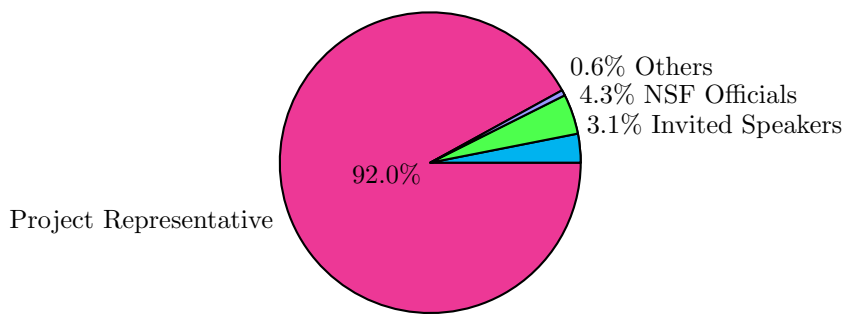
\begin{figure}[h!]
\centering 
    \begin{tikzpicture}
    \pie[radius=2,color={cyan!90, green!70, blue!40, magenta!90}]{3.1/, 4.3/, 0.6/, 92.0/Project Representative}
    \pie[hide number, radius=2, color={cyan!90, green!70, blue!40}]{3.1/\!3.1\% Invited Speakers, 4.3/\!4.3\% NSF Officials, 0.6/\raisebox{3pt}{\!0.6\% Others}}
    \end{tikzpicture}
    \caption{Primary role of survey respondents at the meeting}
    \label{fig-feedback-respondent-meeting-roles}
\end{figure}

\begin{figure}[h!]
\centering
    \begin{tikzpicture}
    \pie[radius=2,sum=auto]{
        86/Faculty with teaching responsibilities,
        7/NSF Official,
        61/Research faculty\, scientist\, staff ,
        8/Other
    }
    \end{tikzpicture}
    \caption{Primary role of survey respondents at their home institution.}
    \label{fig-feedback-respondent-home-roles}
\end{figure}
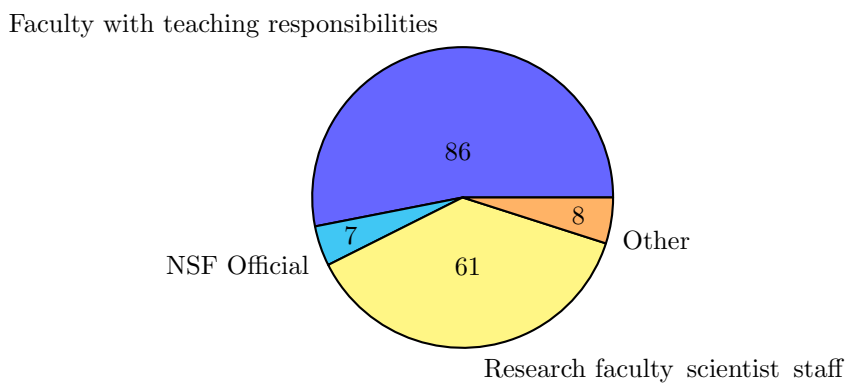

Our respondent's roles at the meetings and roles at their institutions are summarized in Figures \ref{fig-feedback-respondent-meeting-roles} and \ref{fig-feedback-respondent-home-roles}.

\subsection{Feedback on Meeting Organization}
About 93\% of respondents were neutral or found the meeting organization positive. From their written feedback on meeting organization, respondents said the short time frame for announcements resulted in issues with planning travel and accommodation, and had conflicted with other personal commitments. Some respondents experienced additional financial costs because of the short timeframe. Otherwise, respondents were generally pleased with the communication about the meeting. 

\subsection{Feedback on Usefulness of the Meeting}

Respondents were asked to provide their disposition towards different components of the PI meeting as either positive, neutral or negative. The results are presented in Table \ref{tab:feedback-disposition}. Respondents were also allowed to provide textual feedback.

From the text feedback, the poster sessions were the most positive experience of the PI meeting with some respondents suggesting that there be more poster sessions and that they be held for longer periods of time to allow for more interaction and discussion.
Many respondents highlighted the social events and networking opportunities as critical parts of their experience as it gave them opportunities to meet colleagues across disciplines for future collaboration. The breakout sessions received mixed feedback: respondents wanted more time and more structure for the breakout sessions. However, as we see in  section \ref{sec:breakout} many constructive ideas were contributed in the breakouts for the future development of the program.

\begin{table}[ht!]
    \centering
    \begin{tabular}{lccc}\toprule
    \textbf{Event Category} & \textbf{Positive} & \textbf{Neutral} & \textbf{Negative} \\ \midrule 
     Poster Sessions  & 147 & 16 & 1 \\
     Social Events    & 134 & 28 & 2 \\
     Networking / Informal Engagements & 128 & 35 & 1 \\
     NSF Program Highlights & 113 & 43 & 8 \\
     Panels & 89 & 61 & 14 \\ 
     Breakout Sessions & 71 & 64 & 29 \\ 
     NSF Office Hours & 65 & 91 & 8 \\ \bottomrule
    \end{tabular}
    \caption{Disposition of respondents towards different components of the PI meeting.}
    \label{tab:feedback-disposition}
\end{table}

\subsection{Feedback on Meeting Program and Events}
Most attendees were pleased with the format of the meeting and did not express any specific changes. 
The remaining attendees' feedback is summarized as follows.
\begin{enumerate}
    \item Attendees wanted more time for interaction, including opportunities for structured networking, sessions to interact with other PIs on best practices and potential for collaboration, and more time during poster sessions.
    \item Attendees wanted more time for NSF office hours as only a few slots were available per program officer in attendance.
    \item Attendees were not as engaged with panel sessions and preferred more targeted discussions in smaller groups.
    \item Attendees found the breakout sessions too broad in focus and too short in time. They wanted clearer goals for the sessions.
    \item Attendees found the focus on AI/ML too strong in panels and breakouts, leaving them feeling left out if their topic did not coincide with AI/ML.
    \item Some attendees suggested an event specific to first time NSF PI Meeting attendees.
\end{enumerate}

\subsection{Feedback on Meeting Location and Timing}

Most attendees were pleased with the location with the meeting in Charlotte, NC as it was a central hub for air travel. However, they found the venue to be isolated from nearby restaurants and cafes, making it difficult for them to socialize after the meeting period. 
Many attendees suggested Washington D.C., Chicago and Denver as desirable meeting locations as these places are easily accessible or centrally located in the country.

The timing of the meeting for arriving on a Sunday and departing on a Tuesday were generally positively viewed. Respondents suggested that August 10 to August 12 of next year would be an ideal time to have the meeting.

\subsection{Feedback on the Costs of Attending the Meeting}

Respondents were split on the registration fee: some found the fee reasonable, while others expressed dissatisfaction with a registration fee for a mandatory meeting that is funded by the NSF. Respondents also commented on the high hotel rate. 

\subsection{Summary Recommendations Based on Feedback}
\begin{itemize}
    \item {\bf Location:} should be central and easily accessible (e.g.  Washington D.C., Chicago and Denver) with affordable hotel rates.
    \item{\bf Timing:} Mid-August Sunday-Tuesday timing was satisfactory though earlier notice will be helpful for planning travel. 
    \item{\bf Additional time for discussion and networking:} PIs expressed need for additional time for networking and structured/unstructured time to learn best practices.
    \item {\bf Panels and Breakouts:} Panels attracted poor response while the breakouts were felt to be useful but rushed and not fine grained enough. Recommend reducing panels and using the time saved for more focused breakouts. 
    \item{\bf NSF PD Office Hrs:} Additional office hours if possible are recommended.
\end{itemize}

\section{Panels and Keynotes}
\label{sec:kenotes}

During the meeting, there was one Keynote Speaker session (on Day 1) and four Panel sessions, spread across Days 1 and 2. Details follow in the sections below.

\subsection{Keynote Speaker Summary}
The Keynote address, titled ``Update on NSF’s Office of Advanced Cyberinfrastructure and the National AI
Research Resource Pilot," was presented by Katie Antypas, Director, Office of Advanced Cyberinfrastructure. A copy of her presentation can be found here: \href{https://drive.google.com/file/d/1xZi-rp7ADXCWZiDlVZmjyQnZltdVYpTH/view?usp=sharing}{[PDF]}

The keynote provided the PIs with a great overview of NSF/OAC current and future investments and thinking for the future. Keynote was well attended and inspired/ informed much discussion among attendees. Such a keynote that is effectively a ``State of the Programs and Division" address is very helpful and serves well to set the tone of the meeting.

\subsection{Panel Session Summaries}  
The panels were designed to address topics relevant to both the CSSI and the CyberTraining/SCIPE communities. They were organized around a topic, and panelists were asked to answer a set of questions. The panel sessions were then followed by the Breakout sessions (Section \ref{sec:breakout}), where the community met in several small groups and worked together to answer questions related to the panel topics.

Two of the panel sessions were jointly attended by both the CSSI and CyberTraining/SCIPE participants, and one session was dedicated to the separate NSF programs. The Panel sessions are listed in Table \ref{tab:panels}.

\begin{table}[h]
\label{tab:panels}
\vspace{6pt}
\caption{Panel session details}
\noindent\makebox[\textwidth]{
\begin{tabular}{|p{1.5in} | p{1.8in} | p{1.0in} | p{2.2in} |} 
\toprule
\textbf{Panel} & \textbf{Topic} & \textbf{Moderator} & \textbf{Panelists} \\
\hline
Panel I: Joint & CI for AI In Science: Role of NAIRR & Dan Stanzione, (TACC/UT Austin) & Katie Antypas (NSF), Varun Chandola(NSF), Ben Brown (DOE/OASCR), D. K. Panda(OSU) \\
\hline
Panel IIa: CSSI &  Future-Proofing Investments for CSSI Tools: 	Sustaining tools over changing hardware and software landscapes & Abani Patra, (Tufts U.) & Ian Foster (U. Chicago/ Argonne), Dan Negrut (UW Madison), Daniel Crawford(Virgina Tech), Mike Heroux(Sandia) \\
\hline
Panel IIb: CyTr/SCIPE & Community experiences and Evolving Needs & Alan Sussman (U. Maryland) & Manish Parashar (Utah), David Hart (NCAR), Ritu Arora (Wayne State), Mary Ann Leung (Sustainable Horizons Institute) \\
\hline
Panel III: Joint & The People Resource Gap & Jeff Carver (U. Alabama) & Dan Katz (UIUC), Sandra Gesing (US-RSE), Michael Zentner (UCSD/SDSC), Chuang Wang (UNCC) \\
\bottomrule
\end{tabular}
}
\end{table}

\subsubsection{Panel I (Joint): CI for AI In Science: Role of NAIRR}
\textbf{Moderator:}  Dan Stanzione, (TACC/UT Austin) \\
\textbf{Panelists:} Katie Antypas (NSF), Varun
Chandola(NSF), Ben Brown (DOE/OASCR), D. K. Panda(OSU) \\
This panel  explored challenges and opportunities in provisioning CI for the AI revolution in scientific discovery and engineering innovation. AI for science requires CI comprised of hardware, software, data and people resources acting in concert. Core questions   addressed included:
\begin{enumerate}
\item What are the major gaps in current CI available to support AI driven science? In particular, what aspects of existing CI designed for more traditional compute intensive sciences need to be expanded. 
\item What are challenges in training and preparing the workforce needed?
\item What are the current resources and opportunities for investigators?
\end{enumerate}

\subsubsection{Panel IIa (CSSI): Future-Proofing Investments for CSSI Tools}
\textbf{Moderator:} Abani Patra (Tufts University)\\
\textbf{Panelists:} Dan Katz (UIUC), Sandra Gesing
(US-RSE), Michael Zentner (UCSD/SDSC), Chuang Wang (UNCC)\\
\textbf{Questions for Discussion}:
\begin{itemize}
\item Sustaining tools over changing hardware and software landscapes
\item Collaboration between domain scientists and computer/ software engineers
\item HPC Resources are shifting towards GPUs, TensorCores, etc. How do we cope?
\end{itemize}

\subsubsection{Panel IIb (CyTr/SCIPE): Community experiences and Evolving Needs }
\textbf{Moderator:} Alan Sussman \\
\textbf{Panelists:}  Manish Parashar (Utah), David Hart (NCAR), Ritu Arora (Wayne State), Mary Ann Leung (Sustainable Horizons Institute)\\
\textbf{Questions for Discussion}:
\begin{itemize}
\item Metrics for Cybertraining/SCIPE programs and outreach
\item Large Institutes - how do they get created? 
\item Community Building/Broader Impacts:
\begin{itemize}
\item Create broader communities, reach out to MSIs; grow new PIs
\end{itemize}
\item Promoting Better Scientific Software
\item New Ways/How to multiply the NSF investment – delivering diverse courses, certificates, programs, different domains
\item How to effectively broaden the science/engineering research workforce, beyond a one size fits all approach
\end{itemize}

\subsubsection{Panel III (Joint): The People Resource Gap }

\textbf{Moderator:} Jeff Carver \\
\textbf{Panelists:} Dan Katz (UIUC), Sandra Gesing (US-RSE), Michael Zentner (UCSD/SDSC), Chuang Wang (UNCC) \\
\textbf{Topic:} The People Resource Gap - RSEs, Data/AI scientists \\
\textbf{Discussion Topics}:
\begin{itemize}
    \item Topic 1 - Pipeline
    \begin{itemize}
        \item Driving Question: In your opinion, how should we design training programs to draw people into RSE and related careers? Including:
        \begin{itemize}
            \item Training on specific skills (including software engineering and other skills)
            \item Training programs that integrate multiple skills into a coherent program
            \item Structure of training programs (e.g. short courses, apprenticeships)
            \item Format of training programs (e.g. online, in-person, asynchronous video)
            \item Credentials
        \end{itemize}
    \end{itemize}
    \item Topic 2 - Professional Development
    \begin{itemize}
        \item Driving Question: How can organizations best support the career development of people in RSE and other related roles? Including:
        \begin{itemize}
            \item What types of professional development do they need?
            \item Are these roles careers in themselves or a step along the way to something else?
            \item How can people in these roles advance within their organizations?
            \item What are the impediments to advancement?
        \end{itemize}
    \end{itemize}
\end{itemize}


\section{Breakout Session Summaries   } 
\label{sec:breakout}
Breakout sessions following keynotes provided a key forum for seeking input from the attending PIs and project representatives. Each breakout was moderated and key questions were posed to start discussions. Notes taken at the sessions formed the basis for the session summaries below.  Where applicable, summaries include the following sections: Overview; Current Status/ Challenges; Desired Outcomes and Future Directions; Recommendations.

Original notes can be found in the meeting google drive Breakout Session folder: 
\url{https://drive.google.com/drive/u/0/folders/1SUKTkqF5I3_gHl7jR_pM8KMpdO64cELH}

\subsection{Enabling the ML/AI Transformation of Science Discovery and Innovation }
\label{sec:breakout_1}

Embedding machine learning (ML) and artificial intelligence (AI) methods into scientific tools presents a range of challenges and opportunities. Breakout participants have provided thoughtful input to define primary challenges, desired outcomes and recommendations to enable this transformation.
\subsubsection{Current Status/ Challenges}
\begin{itemize}
\item  At the heart of these challenges is the diversity and complexity of scientific data. Scientific datasets are often heterogeneous, coming from various sources and in different formats, which complicates their integration and processing for ML/AI models. 

\item Additionally, high-quality, labeled data, which is crucial for effective ML/AI training, is frequently incomplete or noisy   in scientific fields.

\item Another significant challenge is scalability. Fields like genomics and climate modeling generate massive datasets that require substantial computational resources and sophisticated algorithms to manage effectively. 

\item This issue is compounded by a lack of interdisciplinary expertise; many domain scientists are not well-versed in ML/AI, which limits their ability to leverage these technologies fully. 

\item Moreover, the transferability of ML models across different disciplines is problematic, as best practices and models can vary widely, making it difficult to select the most appropriate approach for specific scientific problems.

\item The validation and interpretability of ML/AI results are also critical concerns. For ML/AI to be useful in scientific research, results must be validated and provide interpretable insights into fundamental science questions. 

\item Additionally, infrastructure and standardization issues further complicate the integration of ML/AI into scientific workflows. The difficulty in switching between different AI tools, such as PyTorch and TensorFlow, hampers collaboration and adoption, while the lack of standardized frameworks and processes affects consistency and usability.

\item Preparing data for ML/AI applications adds another layer of complexity. The preprocessing of data can be labor-intensive and time-consuming, with issues related to data quality and metadata complicating the process further.
\end{itemize}

\subsubsection{Desired Outcomes / Future Directions}
\label{sec:sec:breakout_1-fut}
A primary desired outcome is a skilled and well trained workforce capable of enabling and sustaining the ML/AI transformation. Therefore, investing in enhanced training and education for domain scientists is crucial. In the future we there will be available training on ML/AI methodologies for both specialists and domain scientists and accessible workflows and infrastructure  to significantly improve our ability to utilize these technologies. Additionally, support for integrating ML-based courses into graduate programs can help build foundational knowledge and skills.

Low barrier access to high end CI (computing, data, workflows and training harnesses) that allows rapid development and exploration will maximize science advances and innovation. Thus, 
improving infrastructure and resources is a  key priority. Leveraging initiatives like the National AI Resource Research (NAIRR) can provide access to necessary computing resources and promote the standardization of AI tools and models. Developing AI consulting services similar to HPC consulting could also help researchers integrate ML/AI best practices into their work.

Solutions in place for data management and interoperability   for effective ML/AI integration. Promoting standardized data formats and frameworks can enhance data quality and facilitate interoperability between different AI tools. Improving data pipelines for preprocessing and integration will make data more accessible and useful for ML/AI models. Investing in hybrid solutions that combine AI with traditional computational methods can help tackle domain-specific challenges more effectively.

Fostering community and collaboration is vital for advancing ML/AI in scientific research. Encouraging cross-disciplinary collaboration between computer scientists and domain experts can lead to the development of domain-specific AI solutions and methodologies. Creating benchmarking metrics and evaluation datasets tailored to scientific domains will also help in validating and comparing AI models.

Sustainability and long-term investments are crucial for maintaining progress in AI research. Addressing the need for long-term funding for personnel and developing career paths for engineers and developers are important steps. 

Lastly, ensuring that AI resources are inclusive and accessible to a diverse range of researchers can enhance the utility of AI approaches. By addressing these challenges and leveraging the identified opportunities, CSSI and related communities can better integrate ML/AI methods into tools for scientific discovery and innovation.

\subsubsection{Recommendations}
\label{sec:breakout_1-rec}
The following recommendations summarize the outcomes of the discussion:
\begin{itemize}
    \item {\bf Investments leading to low barrier access to high end CI} (computing, data, benchmarks,  workflows and training harnesses) to enable rapid development and exploration will maximize science advances and innovation.
    \item {\bf Support  innovative approaches to training on ML/AI methodologies, workflow development, and infrastructure usage.} Such training in multiple modalities -- workshops to classes for both domain scientists and specialized CI personnel like research software engineers on a priority basis at scale where adequate numbers of personnel at every major research institution has access to such training.
    \item {\bf Investments in  standardized data formats and robust frameworks }to enhance data quality and facilitate interoperability between different AI tools.
    \item {\bf Develop programs that address the need for long-term support for personnel} and  career paths for engineers and developers who enable the development of AI/ML resources.
    \item {\bf Ensure that AI resources are inclusive and accessible to a diverse range of researchers }from different geographic regions and demographics.

\end{itemize}
\label{sec:breakout-enabling_MLAI}

\subsection{Access to Hardware Resources }
\label{sec:breakout_2}


\subsubsection{Background}

We opened with a discussion of what we mean by hardware when it comes to CI. Hardware includes traditional CI resources (e.g. high-performance computing); however,  In the modern AI era, the notion of hardware is augmented with a heterogeneous mix of specialized accelerators (FPGAs, GPGPUs [of course], chiplet designs, and other novel systems found only in specialized labs (e.g. Cerebras, Graphcore, SambaNova, others). Beyond computation, storage remains a challenge, where hosting a multiple TB or PB scale dataset remains a challenge, including reliable long-term backup and permanent hosting/dissemination of large datasets to enable reproducible science.  Beyond hardware, there are the ongoing costs to keep the systems running (maintenance, system/network admins) and to help users (research engineers/RSEs).

\subsubsection{Current Status/Challenges}

The following represent key pain points/challenges when it comes to hardware to support CI research:

\begin{itemize}
\item Data Storage and Migration: The inefficiencies and time waste associated with migrating data across different locations are significant impediments.

\item Co-Located Compute and Storage: The lack of co-located compute resources with storage creates bottlenecks in processing efficiency.

\item Software Compatibility and Interoperability: Heterogeneous computing introduces software incompatibility issues, necessitating better support for interoperability frameworks.

\item Training and Personnel: There is a clear need for more training resources and skilled personnel to support these systems.

\item Awareness and Access to Resources: Researchers are often unaware of available CI resources, and existing portals do not make access sufficiently user-friendly.

\item Scalable Storage Solutions: AI's storage needs are vast, requiring more scalable solutions that include long-term availability and artifact preservation.

\item Energy Efficiency and Long-Term Costs: As hardware obsolescence slows, energy efficiency and the long-term costs of running AI hardware become critical considerations when determining whether a resource is obsolete and is worth the ongoing power and maintenance costs.

\item Flexible and Sustained Funding Models: There is a need for more flexible and sustained funding models that support both short-term maintenance and long-term hardware investments.

\item Cloud and On-Premises Balance: While cloud computing offers flexibility, it can be costly for experimentation, highlighting the need for a balanced approach between cloud and on-premises resources. Nevertheless, any on-premises or national CI resources should offer a compelling option to what is possible in the cloud as many researchers are opting for commercial solutions.
\end{itemize}

\subsubsection{Desired Outcomes / Future Directions}
\label{sec:breakout_2-fut}

If the recommended changes were implemented, several scientific use cases would flourish, particularly in fields requiring large-scale data processing and long-term storage. Environmental modeling, genomic research, and AI-driven simulations in physics are prime examples where access to flexible, scalable storage solutions and high-performance computing resources is critical. These projects often generate massive datasets that need to be processed efficiently and stored securely over extended periods, something that academic infrastructure can provide more effectively than commercial clouds. The flexibility of academic environments allows for customized hardware setups, such as heterogeneous computing platforms and the ``condo model" for shared resources, which can be optimized for the specific and evolving needs of these complex research endeavors.

Beyond just cost and efficiency, this flexibility fosters creativity in how researchers approach their work. Sensitive research areas like genomics or social science studies benefit from the ability to implement stringent data security measures and develop long-term data management strategies within dedicated academic data centers, rather than relying on the more rigid frameworks of commercial clouds. Moreover, the freedom to tailor computing environments to meet the unique demands of interdisciplinary AI research or unconventional data processing enables researchers to explore new methodologies and push the boundaries of their fields. By investing in these adaptable and creative infrastructure solutions, NSF-funded projects can avoid the limitations of commercial cloud dependency, such as unpredictable costs and challenges with long-term data accessibility, while enhancing the innovative potential of scientific research.

\subsubsection{Recommendations}
\label{sec:breakout_2-rec}
We propose the following short-term recommendations:

\begin{itemize}
\item \textbf{Introduce Small, EAGER-like Awards for Hardware Maintenance and Upgrades:} Create a mechanism within NSF for small, targeted awards that allow for the maintenance and incremental upgrades of existing hardware throughout a project's lifecycle (even beyond the grant period). This would help ensure that equipment remains functional and up-to-date, minimizing disruptions between awards and performance issues. 

\item \textbf{Enhance Support for Scalable Data Storage Solutions:} Provide supplemental funding specifically aimed at high-capacity data storage solutions. This could involve integrating on-premises storage with cloud-based systems to better meet the large-scale demands of AI-driven research projects.

\item \textbf{Expand Training Programs for Heterogeneous Computing and Interoperability:} Increase funding for workshops and training programs focused on heterogeneous computing environments and software interoperability. These programs should target both students and researchers, equipping them with the skills needed to effectively manage and utilize diverse hardware and software systems.

\item \textbf{Develop Campus-Level "Condo Model" for Shared Hardware Resources:} Offer grants that support the adoption of a "condo model" for shared hardware resources on campuses. In this model, centralized resources are managed at the institutional level, while individual research groups have the option to add dedicated nodes, optimizing resource use and reducing costs.

\item \textbf{Fund Studies on Improving AI Hardware Energy Efficiency:} Provide funding for empirical  studies and pilot projects that focus on retrofitting existing AI hardware to improve energy efficiency. This would help reduce the long-term operational costs associated with running AI-driven research, making it more sustainable.
\end{itemize}

We propose the following long-term recommendations:

\begin{itemize}
\item \textbf{Establish a Long-Term Hardware Support Fund:} Develop a dedicated fund within NSF grants that supports the full lifecycle of hardware, from acquisition to decommissioning. This should include provisions for ongoing maintenance and energy efficiency improvements, ensuring long-term usability and sustainability of research infrastructure.
\item \textbf{Develop a Sustainable Data Management Strategy:} Implement a long-term strategy for data storage and management that includes the preservation of research artifacts and datasets. This could involve the creation of NSF-funded data centers dedicated to ensuring that data generated by NSF-funded projects remains accessible and usable over time. Leverage what other agencies know about managing extreme data, e.g. data storage from high-energy physics (DOE), and ingesting all internet traffic (NSA).
\item \textbf{Establish Permanent NSF-Funded Training Centers:} Create NSF-funded training centers that offer ongoing support and education in heterogeneous computing and software engineering/interoperability. These centers could also serve as hubs for the development and testing of new frameworks, ensuring that researchers are equipped with the latest tools and knowledge.

\item \textbf{Promote Regional or National "Condo Models" for Shared Resources:} Encourage the development of regional or national "condo models" for hardware resources, where multiple institutions share access to high-performance computing resources. This would reduce costs per researcher and increase access to state-of-the-art technology across the academic community.

\item \textbf{Prioritize Energy Efficiency in Hardware Funding Decisions:} Make energy efficiency a key criterion in NSF’s hardware funding decisions. By prioritizing equipment and infrastructure that offer better energy performance, the NSF can help ensure that AI research remains sustainable both financially and environmentally.

\end{itemize}

\label{sec:breakout-access-hardware}

\subsection{Role of RSEs and Other Support Staff  }
\label{sec:breakout-rse}
\label{sec:breakout_3}

\subsubsection{Background}
This topic focuses on the roles that various software-facing CI Professionals (i.e. Research Software Engineers, Data Scientists, and Research Infrastructure Engineers) play in the research ecosystem, how those roles may evolve in the future and what support the community needs to facilitate these changes.

\subsubsection{Current Status}
Across domains and across institutions, it is clear that there are a wide range of roles that RSEs play in STEM research \& education, both in terms of the kinds of work that they do and the formal positions they hold. Examples included:

\begin{itemize}
\item Faculty/PIs and graduate students who need computational tools or novel software for their scientific work, self-teach or train for the skills needed, and create purpose-built software on their own.
\item Research labs where a lab member becomes the primary specialist in software and coding skills, serving as a resource either formally (a software engineer or other technical expertise, hired for software support) or informally (a graduate student or postdoc who specializes and teaches others).
\item Full-time RSEs who work full-time, funded as staff on a single grant, often housed in an academic department, but sometimes a university’s central RSE institute. These staff may have PhDs (or otherwise have deep domain specialist knowledge), coupled with more informal/on-the-job software skills. Or, they may have a more traditional software background and then learn to work with scientific collaborators.
\item Software infrastructure engineers that are operationally focused via software development in support of research, education, and instruction. These are more like modern IT professionals in that they have application services that they maintain, defined in software, that are related to networking, security, identity, and cloudy infrastructure.
\item Attendees from NCAR \& UCSD represented national centers with full-time RSEs who work on rotating portfolios of projects with scientists. 
\end{itemize}

Across contexts, many attendees spoke about the challenges of finding and retaining staff for these roles, which require project-specific combinations of software engineering, scientific, and communication/project management skills. People in these roles are often faced with communication barriers between researchers and engineers, who are largely trained to describe projects in quite different terms.  Engineers are most effective if they are given clear requirements but researchers don’t want to commit the time to developing them (and suffer negative career consequences if they do). Software engineering teams are seen as current “high value/high impact” for research-facing activities in academia, but are also the hardest to retain.

\subsubsection{Desired Outcomes / Future Directions}
\label{sec:breakout_3-fut}

Discussion groups spoke extensively about the need to provide formal support for RSE time and expertise, and to develop career models that make sense for various contexts, whether that be dedicated career tracks focused on computational work and software engineering, recognizing faculty and research staff efforts, or working with university-level centers such as libraries and computing centers. These roles need to be elevated from “support” to coequal to research staff, closely integrated and integral to making the science happen - through both career recognition, citations, as well as stable career paths, in order to recruit/retain these professionals away from industry. Some important features might include ensuring recognition of software projects as first-class scholarly outputs, and some level of autonomy for RSEs to pursue their own funded research.

Another thread of discussion focused on the need to spend the time to understand what kinds of software expertise and projects are needed where, and what kinds of resources provide the most benefit to different communities of researchers. Differentiating between software applications/workflows and software-as-infrastructure is also important. One attendee summarized: “The needs for specialized services become more clear/focused at departmental scale, this is where RSE’s and other CI professionals are the most impactful for individual science domains.” Others mentioned that many emerging research institutions or departments may need dedicated attention to fundamental data engineering and data management capacities before getting to more complex or high-compute software workflows.

\subsubsection{Recommendations}
\label{sec:breakout_3-rec}
\paragraph{Diverse funding models for stable support }Attendees recognized a need for a variety of different funding models, combining federal grants, institutional funding, and potential fee-for-service and cost-sharing models (e.g. NSF funds first 3 years, university funds after that). For sustainability, granting agencies might need to consider both larger grants for building out significant new capabilities, as well as determining what ongoing needs are required - just as with other physical infrastructure, software infrastructure entails maintenance and ongoing support for users. The shape and scope of collaborations also needs to be able to vary - in many cases, critical software infrastructure is maintained by a small group of people at one or two institutions, but used by a community of scientists from many different institutions. 

\paragraph{Invest in RSEs integral to major CI investments}Both universities and funding agencies are encouraged to invest in RSE staff as a key component in the research enterprise, just as they do with computing or with large hardware investments like MRI, telescope facilities, etc.

\paragraph{Match RSE expertise and experience to domain science} Discussion also focused on maintaining appropriate recognition for the diversity of contexts and needs that software-facing CI professionals operate across. The field needs support for the work it takes to find the right matches between RSE and projects where DE is useful, and for researchers to take the leap of learning to communicate with software engineers.  At the department and domain level, the kinds of interactions that are most important may vary - the resources and projects needed probably will not look the same in chemistry as in political science.  

\paragraph{Support teams of RSEs with different levels and domains of expertise} Different domains and different institutions will also involve a continuum of maturity or professionalization of RSE work. In some cases, research institutions need foundational Data Engineering professionals (management, transport, staging, curation), who may be housed in libraries, software carpentry type training resources, or dedicated software centers, etc. -  before specialized software can be effectively used. In others, teams that combine RSE skills with domain expertise can be embedded in departments or research institutes to serve portfolios of projects that benefit their fields.

\subsection{Designing Specific AI Tools for Science Discovery and Innovation  }
\label{sec:breakout-ai-tools}
\label{sec:breakout_4}

\subsubsection{Background}
Artificial Intelligence (AI) has come to the forefront as an area of national research focus. AI holds promise to aid and even transform science discovery and innovation, yet AI is still in its infancy. How AI tools can be used, which ones should be developed, and how they should be deployed is an evolving preoccupation of the research community. 

\subsubsection{Current Status}

Current AI use cases suggest insight into these questions.
For instance, AI is being used to predict dangerous algal blooms in freshwater nearly two weeks prior to their appearance based on heterogeneous data from solar radiation measurements, water oxygenation levels, precipitation records and nutrient loads in freshwater. AI is also being used to improve parameter exploration in high-fidelity fluid dynamics simulations for a variety of atmospheric and geophysics applications. 

\subsubsection{Desired Outcomes / Future Directions}
\label{sec:breakout_4-fut}

While such applications of AI underscore its potential utility, there are many areas of science discovery and innovation where AI tools have yet to make a substantial impact. 

\paragraph{Potential Use Case 1.} Binary exploitation uses a computer’s standard operations, often memory control behaviors, to gain access to the computer’s data or processes. Often, binary exploitation comes with delivering a payload to the target system to encourage certain types of behavior. The development of heterogeneous payloads such as text and images is an area that is ripe for AI tools.

\paragraph{Potential Use Case 2.} Scientific workflows often require working with a variety of structured and unstructured materials ranging from published documents, metadata and hand-written lab notebooks. As research challenges become more complex, these scientific workflows also become increasingly complex and often distributed amongst multiple scientists and researchers. To support efficient, rigorous collaboration, scientific workflows can benefit from AI tools that can amalgamate, search through, relate and provide access to a variety of relevant research materials. 

\subsubsection{Recommendations}
\label{sec:breakout_4-rec}
To realize these potential opportunities for AI tools in science discovery and innovation, a number of coordinated efforts and investments are needed. 

\paragraph{Early career AI training opportunities} From a workforce development perspective, students need early training in AI/ML tools, including their development and utilization; and individuals need to be encouraged to pursue specialized careers in AI/ML, computer science and domain fields with stable financial support. 

\paragraph{Support for Shared data and sharing mechanisms} From a scientific rigor and democratization perspective, researchers need to be incentivized to develop more shared data and shared cyber-infrastructure resources; researchers need mechanisms to facilitate agreements with private groups and government agencies around data use and harmonization; and researchers need support and training for working with private or high-security data sets. 

\paragraph{Infrastructure support for domain specific AI tools} From a coordination perspective, the community needs federal-level guidance on priorities; support for archiving, accessing and using data and infrastructure after project lifecycles;  and investments into cross-disciplinary efforts to develop AI tools for specific domain problems.

\subsection{Panel Topic Breakout Session: Training Resources for PIs (Domain Scientists) and RSE/Computer Scientists  }
\label{sec:breakout-panel-train-resources}
\label{sec:breakout_5}


\paragraph{Description}  
This panel  explored challenges and opportunities in provisioning CI for the AI revolution in scientific discovery and engineering innovation. AI for science requires CI comprised of hardware, software, data and people resources acting in concert. 

\paragraph{Objective}  
The goal of this breakout is to explore the need for (if it exists), and the challenges and opportunities for developing AI/NAIRR relevant training resources for PIs (domain scientists), RSEs, and computer scientists.  

\subsubsection{Discussion Summary}
In this session, we were asked to answer three questions:

\paragraph{How do we define the NAIRR Workforce? 
Including PIs (domain scientists), RSEs, and computer scientists, more?} 
The group came up with the following list of job categories or titles: Educators, Engineers, Facilitators, Faculty, Master, Post-docs, Professors, Staff, Undergrads, Users	Academic, Domain-science, Experts, Multi-disciplinary, Pipeline, Policy, Teaching.

We then created a list of descriptions for these jobs:
\begin{itemize}
    \item Research scientist (faculty, scientist, post-docs, grad students) for target domains.
    \item Research software engineers as facilitators of technology.
    \item Teaching faculty, educators, and librarians for training and educational resources.
    \item One person may have different hats.
    \item Domain experts that need computer or data scientist support and knowledge.
    \item Domain experts that need to use/learn AI.
    \item Research computer scientists that work with domain experts (mainly grad students). Domain experts know how to compute pre-HPC (e.g. Office applications on a laptop), and need help to use HPC with AI. Research computer scientists need with the transition.
    \item Need research software engineers that are experts in computer science and familiar with the target domains.
    \item It is hard to keep trained personnel without competitive salaries as compared to industry.
    \item Still need high-performance computing experts.
    \item Teaching faculty have the potential to be trained on how to train the research scientist and research software engineers.
    \item Traditional RCD positions 
    \item Data scientists together with domain scientists
    \item Responsible AI, ethics, data wrangler, and data governance 
    \item Legal consultant - acquisition from entities, security
    \item Policy makers 
    \item Evaluations (for different parts of the pipeline / workflows)
    \item We focussed on the academic aspect of the workforce, as that is what we are familiar with. 
    \item The  workforce spans undergraduate students, who are not experts in the domain-science, but, who are nimble at learning software tools and can provide maintenance for installing needed packages
    \item more advanced undergrads/master students who can research best existing software tools and guide undergrads
    \item Postdocs (also staff and professors) who are multi-disciplinary experts in CS / ML-AI / domain science and need to understand
    \item researchers, perhaps using AI tools. 
    \item domain scientists, computer scientists, software engineers 
users of resources such as GPUs/FPGAs
    \item trainers, co-ordinators
    \item pipeline of students/postdocs
    \item policy makers defining the use of data

\end{itemize}

\paragraph{What are the challenges for developing AI/NAIRR relevant training resources for PIs (domain scientists), RSEs, and computer scientists?}
\begin{itemize}
    \item Overcoming silos in goal setting (performance vs actual research goals.), remove barriers for interdisciplinary goals.
    \item Scalability with goals, students don't have time to optimize code, better to run something now than optimize later. Need RSEs knowledge, importance of communication between RSEs and domain experts.
    \item Hard for domain scientists to learn the limitations of the AI ``black-box" tools, and bring previous experience to new workflows.
    \item Hard to know why a AI model worked, "explain" why the model produced the result that it did. Amplified with LLM, how can it be trusted?
    \item Interdisciplinary training.  Bridging between the domain science the data science / AI
    \item Defining the training needs based on a new science
    \item Consolidate tools - easy to use / easy to access
    \item Providing relevant data
    \item Which model/algorithm to apply to a specific application - better filter mechanism of tools
    \item Mismatch in context: what level of fake rate is acceptable.
    \item Excited students might see AI as a silver bullet, and may not understand training data (eg. molecular dynamics simulations) may be very expensive.
    \item As a domain expert, the most difficult issue is talking a different language from computer science. Experience solving it with a bootcamp between RSEs and domain scientists to find a common language.
    \item Really steep learning curve. Harder to train because it is new.
    \item Hard to switch tools once a tool is 
    \item Need knowledge of the target domain.
    \item Access to the scalable AI Infrastructure
    \item Proprietary hardware - Portability 
    \item Language barriers - Data Scientist use one language / domain uses another language
    \item Pre-train or foundation models 
    \item Lack of policies for defining where and when to use AI 
    \item Rapid turn-over in technologies (software \& hardware).
    \item Students are excited about AI, but don’t know which kinds of models are appropriate, how to apply them, data needs, etc.
    \item Language/semantic barrier. Need someone to translate, including understanding which tool might be useable vs one that can not meet target precision needs
    \item AI is multidisciplinary; teaching AI reaches beyond the students’ domain more.
    \item we need to target different audiences/groups; not one size fits all
    \item most of the data, resources, and algorithms are currently proprietary. 
    \item Reliance on commercial models for training materials
    \item technology is moving too quickly. Things are becoming obsolete in a matter of months.
    \item Update existing training material/courses to be current.
    \item Should PhD students be focussed on domain or AI tools?
    \item Retention issues of well trained students/researchers to industry.
    \item Domains such a social science/public policy/medicine have not usually used AI techniques, computer techniques, so there is a large gap.
\end{itemize}

\paragraph{What are the opportunities for developing AI/NAIRR relevant training resources for PIs (domain scientists), RSEs, and computer scientists?} 
\begin{itemize}
    \item Bootcamp to develop common languages, interdisciplinary teams. Bootcamp a good framework to bring people from different domains.
    \item Different domains may want to collaborate in creating the training resources. The same problems are solved over and over again, and this only gets discovered when talking between disciplines.
    \item Make resources searchable, ask for support and collaborate with librarians that know how to organize,and make resources discoverable.
    \item NSF TIP Directorate framework, partner with industry to create knowledge hubs.
    \item Different workshops targeting PIs, domain scientists, RSEs, and computer scientists to get novel work and get the AI trends
    \item Many of the software tools can be taught generically, and not needed by local PIs/postdocs
    \item partnership with industry. Students will be used to using such tools. Industry would like to lock-in students. 
    \item Open source community at the university level.
    \item We could add domain-specific list of tools useful for a given domain, explanation tools, eg. that might recommend something useful for visualization
    \item Webinars that can showcase solutions found, short-videos for a particular skill. The challenge is to break it down in digestible chunks, but still discoverable and relevant.
    \item Similar efforts for High Performance Computing have been accomplished.
    \item Delivering of training - look to non-traditional ways that training is deployed
    \item Partnering with people that use AI
    \item Inclusive training - having a robust catalog of training available for the community
    \item Younger levels in education (AI) - high schoolers, to provide teaching opportunities for them - fundamental AI
    \item Collaborative meetings between domain scientists, computer scientists, RSEs to develop unified AI tools.
    \item Funding agencies seem to be aware for the need of domain-specific RSE funding
    \item We can provide a curated list - that is well known and well communicated to the community - of good AND bad, online learning tools for various tools
    \item We could add domain-specific list of tools useful for a given domain, explanation tools, eg. that might recommend something useful for visualization
    \item If Open Science requirements are made more stringent by funding agencies, eg. required publishing of software, installation instructions etc., it would provide a searchable mechanism within domain-specific areas how others can learn more quickly good tools to use
    \item They could be mandated to do this? But this could have its risks.
\end{itemize}


\subsubsection{Recommendations}
\label{sec:breakout_5-rec}
\begin{itemize}
    \item Train research computer scientists to work with domain experts (mainly grad students). Domain experts know how to compute pre-HPC (e.g. Office applications on a laptop), and need help to use HPC with AI. 
    \item Train research software engineers to be experts in computer science and familiar with the target domains.
    \item Develop a workforce that spans undergraduate students, who are not experts in the domain-science, but, who are nimble at learning software tools and can provide maintenance for installing needed packages
    \item Teach community to overcome silos in goal setting (performance vs actual research goals.) and set realistic objectives
    \item Train users to identify the limitations of ``black-box" AI: what worked, why, etc
    \item Delivering of training - look to non-traditional ways that training is deployed
    \item Provide a searchable mechanism within domain-specific areas so others can learn more quickly and identify good tools to use 
    \item Organize frequent workshops targeting PIs, domain scientists, RSEs, and computer scientists to learn novel methods and keep up with evolving AI trends
    
\end{itemize}


\subsection{ Community building \& measurable broader impacts }

\paragraph{Objective}

The objective of this breakout was to discuss ideas for developing communities in a sustainable manner and measure the impacts of their work.
\subsubsection{Discussion Summary}

\paragraph{ Questions} The discussion was seeded with the following questions:
\begin{itemize}

 \item Can metrics be embedded in how we build communities? 

\item How do you typically measure broader impacts in your community?

\item What would you like to do differently in how broader impacts are considered?

\item Are communities better off being integrated with curricular materials?

\item How should one contextualize metrics for broader impact?

\item What are the privacy issues one should consider?
\end{itemize}







\paragraph{Background}
Broader impacts extend from advancing scientific research and societal outcomes. Broader impacts achieved by building communities is a core tenet of the NSF CSSI, Cybertraining and  SCIPE programs. These three programs target different outcomes, and as such build and support communities with different focuses and interests. While this offers several possibilities in advancing Broader Impacts and societal outcomes, meaningfully measuring their outcomes continues to remain a challenge for these programs. 

\paragraph{Current Status}
The National Science Foundation (NSF) expects researchers' work to have broader impacts: the potential to benefit society and contribute to the achievement of specific, desired societal outcomes. Practices promoting broader impacts have come a long way. Perhaps we are approaching a set minimum expectations.

The NSF does not want to be prescriptive about the societal outcomes a project addresses but provides examples of broader impacts across several categories. These include inclusion, STEM education, public engagement, societal well-being, STEM workforce, partnerships, national security, economic competitiveness, infrastructure. As such, the communities in question can take different forms. Among others, the broader impacts in these communities can be achieved via developer communities, research support structures, curricular adoption, and informal training.  
Practices to enhance the broader impacts in a community are typical today. For example, transitioning ad hoc (or informal) training to the classroom and  sharing curricular materials are increasingly considered standard practices. Today, they need to be considered with the same intentionality as science. It is not sufficient to merely mention activities. When considering broader impacts, researchers should consider the objectives, the activities, and the budgetary considerations. 

Researchers are actively collecting data on extended research outcomes such as publications, software developed, students graduating, funding success, and courses developed. We note that while there are public tools, they may be  hard to use because they might expose student names, FERPA  issues.

Depending on their scope and funding, communities can take years to build.It is hard to gauge the success of community development over the lifetime of a shorter grant. Collecting metrics is somewhat easier in software development projects. Smaller programs are not appropriately funded to perform a true evaluation. Some schools have teams that can assist with evaluations and collect metrics. 

\subsubsection{Desired Outcomes / Future Directions}

The programs in this cohort are developing communities that focus on three critical albeit different aspects of advancing computationally enabled research. The CSSI program is developing communities of developers, and researchers who can adopt them in their workflows. The Cybertraining communities focus on the development and adoption of training materials in informal and formal settings. The SCIPE programs is focused on developing communities that can support researchers effectively use computing technologies.  

The communities in this cohort  will have some overlapping needs, support structures, and ambitions. With this in mind, these communities can morph into others. They will further impact other communities downstream. As such, metrics have to consider the impact on secondary and tertiary communities. The community notes that these impacts may come into play after the duration of the original program. A prescriptive  mechanism that tracks the outcomes and objectives of a community could delineate its progres, but runs into the danger of becoming restrictive. 

Reporting on the Broader Impacts is ingrained in most communities. While these three programs try to capture metrics such as demographics, attendance, publications from their communities, there is a need for more enhanced metrics. Participants in these communities engage for different reasons. While some communities are transient, others may exist for longer. One may consider how we can measure the sense of belonging in a community.  Perhaps a mechanism that supports a longitudinal study could capture this information. 
There is an opportunity to elevate the discussion on what we are collecting. For example, there is a deeper need for mechanisms that help researchers develop, adopt and share new scientific practices. Unlike research outcomes, it is harder to capture this need in metrics.  While we are focused on advancing the use of metrics in a community, mechanisms to protect the participating researchers' privacy, and intellectual property need to be considered as well. It is possible that other programs have developed unique ways to address this challenge. Here, specialized expertise for creating and evaluating approaches might prove to be helpful.  A group structured to meaningfully assist researchers engaging in these communities collectively could be helpful.  

\subsubsection{Recommendations}
\paragraph{
Refine and Expand Metrics}

Collaborate with social scientists to develop new metrics that capture both quantitative and qualitative aspects of community engagement. Consider what tools can analyze community sentiment, diversity, and long-term engagement trends. Allow communities to define what matters to them. 

\paragraph{Innovate Broader Impact Strategies}

Encourage original thinking in broader impact planning. Ensure appropriate budget allocations for long-term community engagement and sustainability initiatives. Explore the use of unstructured surveys to gain deeper insights. 

\paragraph{Strengthen Privacy and Legal Compliance}

Work with legal experts to ensure all tools and methods used in community-building are compliant with Federal, State and privacy laws. Consider adopting practices like partial name storage or anonymization to protect personally identifiable information while maintaining functionality.  Community leads must make a habit of requesting consent for all personal information collected and leveraging their institutional review boards (IRBs).
Invest in Expertise
There is a need for a coordinated effort that brings expertise together in a cohesive manner. 
Mechanisms could include hiring or consulting with social scientists or other specialists in com
munity engagement to guide the development of community-building initiatives. Provide training and resources to investigators to help them better understand and implement these strategies.


\subsection{Translating Science Advances into CSSI Tools: from Papers to Software  }
\label{sec:breakout-cssi-tools}
\label{sec:breakout2_2}

\subsubsection{Background}
The CSSI program has traditionally enabled the computational science community to translate new scientific innovations in modeling and related ideas into tools for widespread community usage. \cite{CSSI-Program}
Our group here presented their thoughts on  the principal challenges associated with this goal of CSSI. The session focused on addressing the question 
``What are the main challenges to developing and sustaining new community tools for enabling and representing new science advances?"

\subsubsection{Current Status}
\begin{itemize}
\item   Getting credit and recognition for developing community software and tools continues to be a challenge. Public resources for sharing software like github are  increasingly useful and reduces the problem for students going to industry but academic recognition is still an ongoing challenge. Introduction of new journals like Journal of Open Scientific Software (JOSS) \cite{JOSS} provide mechanisms for getting metrics like citations to support traditional measures of academic success. However, the core problem of recognition remains since many software tools for science are intrinsically harder to generate and such measures of credit limited.
The core workforce of computational and domain scientists need recognition to create a sense of ownership and accomplishment for success. Tool and software development is often perceived as a career killer for science personnel as a consequence of this lack of recognition.

\item Maintenance and sustainability of new tools is really hard – getting money for updating and maintaining software is really hard. 
In most domain getting resources for hiring and supporting Computing Science personnel remains a challenge even in support of big instruments (e.g. LIGO). Support for this staff is critical but not understood and rarely budgeted for though elements of programs like the sustainability track of CSSI are helping. More permanent positions that are not dependent on single tools are needed for engineers who need to build careers. 
\item Furthermore, just keeping software tools functioning is inadequate since new scientific advances need to be continuously assimilated to keep tools scientifically relevant. Developing new tools for minor advances and/or similar models that can benefit from the same programming and data infrastructure is wasteful but often the ``novelty" of the science often makes it easier to find support for such development. Summarizing, a principal challenge to developing tools that translate new computational science into tools that make it seamlessly available to the community remains sustained funding.


\item A persistent challenge is the need for effectively delivering tools to  domain scientists. Well organized communications and community building and outreach through workshops, town halls and such are needed. Platforms like github now provide mechanisms for sharing but community needs to be built and sustained. 
Progress on community specific platforms and science gateways(see for e.g. \cite{nanohub,SGX3})
 with extensive investments through the CSSI programs and domain science specific programs has been very good. Many such platforms are doing well though few have solved the sustainability challenges. \cite{SGX3} designated as a software institute has created much support and programming for science gateways serving  large range of science communities.


\item Software tools especially those that have complex usage modalities may take a long time for uptake. We need to bring community stakeholders and tool developers together - bridging the domain/CI gap. We need to evaluate the future value of the work and sustain it through the uptake cycle. Tools that meet a pressing need for the community tend to be more popular. However, it is not always possible to make a tool flexible enough to meet the needs of the community.   
Thus, generalizing a niche tool used for an advance is important. The tension between specialization (useful in specific context) vs. generalization (usefulness to broader community) is one that each community and tool need to develop.

\item While we started this section highlighting the insufficient acknowledgement / recognition of tool development efforts, insufficient recognition of the interdisciplinary efforts needed here are also often lost.

\item Much scientific software still has a dependence on the student teams that may not last beyond students’ graduation dates for creating tools based on new innovations (often in the student's dissertations). Such tools are often inadequately developed for
scaling to more users. 

\end{itemize}
\subsubsection{Desired Outcomes/ Future Directions}
\label{sec:breakout2_2-fut}
Software developers and other enablers for computing and data driven science should be first class citizens of the research ecosystem. 

The MOLSSI (one of two software institutes that were funded) focuses on computational chemistry tools and ha built a healthy ecosystem of domain scientists, software engineers and support personnel and  trains UG and Grad students to learn to write sustainable code and launch into community. This is an excellent model. 

The desired stable state of processes for effective translation of science advances to tools requires both infrastructure(hardware, tools and people) and culture. A principal part of the culture is a focus on reproducibility in science done with computing. 

Sustainability of tools with a variety of resources to support the needed people and platforms is a much desired goal.
Globus is an example of a widely used, open access software, that is supported via a model that works: freemium with premium features available to subscribing institutions.

  Successful CSSI tools often have long-term champions (e.g. the Department of Energy National Labs support a number of tools -- e.g. Trilinos and PetSc) who bring in funding, personnel and the continued intellectual inputs needed.
  
\subsubsection{Recommendations}
\label{sec:breakout2_2-rec}

\begin{itemize}
    \item {\bf Encourage mechanisms for providing credit for software and tools.} Require DOI’s for publicly available software/data resources; citations and papers in appropriate journals. Encourage inclusion of tools and their usage in grant reporting across both CSSI and domain science grants to promote recognition of software development activity.
    \item {\bf Sustainability grants for software tools for science.} The need grants for sustainability to keep software working was reinforced. Recent addition of the sustainability track was recognized. Such support ``keeps the lights on and provides a base for rapid adoption and sustained use of computing driven science.  
    \item {\bf Emphasize science culture with reproducibility of science advances and reuse of available scientific tools.} Reproducibility initiatives that promote widespread sharing of scientific tools based on science advances help.
    \item {\bf Balanced funding sources.} Certain funding sources should prioritize students while others support research scientists and engineers who can develop and maintain professional quality software.
\end{itemize}


Bridging the gap between community stakeholders and tool developers is crucial for interdisciplinary collaboration. There is often a lack of recognition for these efforts and the challenges of balancing specialized tools with broader generalization. Tools may not be immediately appreciated by the community, and reliance on student teams poses risks if they are not sustained beyond graduation. Funding priorities sometimes favor students over research scientists and engineers, impacting the development and maintenance of high-quality software. Long-term support, potentially modeled after National Labs, is important for scaling and updating software. Globus is highlighted as an example of effective software support through a freemium model that balances open access with premium features.



\subsection{Integrating AI/Foundation Models into CSSI  }
\label{sec:breakout-integ-ai-into-cssi}
\label{sec:breakout2_3}

\subsubsection{Background}
Foundation models have become extremely popular with the public owing to their humorous natural language capabilities and with developers because of their development into programming assistance tools. Owing to these features, determining whether foundation models can benefit cyberinfrastructure for sustained scientific innovation has become an important concern.

\subsubsection{Current Status}
Several existing applications of foundation models for cyberinfrastructure suggest that there is a potential benefit of such models to CI for scientific innovation. 

For example, foundation models for materials science and chemistry learn a representation of the underlying chemistry so that a point in this representation captures relevant physical chemical data of the system. This representation can be used to build other cyberinfrastructure for downstream applications in materials science and chemistry, including simulations and experimental design. 

Foundation models are also used to represent complex, large-scale genomic information. In turn, these foundation models can be used to monitor and predict behaviors of concerning pathogens, such as predicting SARS-CoV-2 variants prior to their emergence. 

\subsubsection{Desired Outcomes / Future Directions}

These successes suggest that foundation models can be part of, spur new, or help build, novel cyberinfrastructure for sustained scientific development. Examples of potential use cases and future directions are listed below, the desired benefit of realizing these use cases, and the challenges to achieving them.

\textbf{Potential Use Case 1.} As hardware continues to evolve and computing systems become more heterogeneous, software often needs to be modified and updated to ensure that they can take advantage of these capabilities to deliver faster or more energy efficient results. However, writing such software is rather challenging and requires developers to not only know how to exploit the hardware, but also requires them to know the scientific domain to ensure the integrity of results. 

Owing to the difficulty of this task, foundation models that can write high-concurrency code for heterogeneous architectures automatically can be very valuable. Such a foundation model would dramatically improve the performance of software and ensure its sustainability even as hardware continues to change. 

Challenges to achieving this use case include developing adequate data for training such models; ensuring that the results are correct; and ensuring that hardware capabilities are being exploited. 

\textbf{Potential Use Case 2.} Cyberinfrastructure for science depends on a clear understanding of the literature in the scientific domain. As manuscript and publication rates grow, CI development needs support for analyzing developments in the literature. 

Developing foundation models for this task can help developers stay on top of advancing research and identify opportunities for improving CI. 

Challenges to achieving this use case include how this data should be collected, processed and referenced by the foundation models, especially in light of copyrighted materials that are mediated by publishers.

\subsubsection{Recommendations}
\label{sec:breakout2_3-rec}

To address the challenges of using foundation models for cyberinfrastructure for sustained scientific innovation, there are two primary recommendations.
First, there needs to be clear domain-specific standards for data collection, data curation and data sharing. The resulting data can be used to train foundation models for use in CI. Second, there needs to be a substantial investment in developing this data infrastructure, and for training individuals to maintain, contribute to, and use this data infrastructure.

\subsection{Integrating New Hardware into CSSI Software }
\label{sec:breakout-integr-hrdwr-into-cssi}
\label{sec:breakout2_4}


\textbf{Objective} \\
How can we more effectively bring CSSI-produced software to the wide group of users the ACCESS/HPC Resource Providers reach?   What are the barriers to integration? 

\textbf{Suggested Questions} \\
See report template below, which can help guide question development. 
\begin{itemize}
    \item Why isn’t your CSSI software product available through ACCESS?  
    \item Should it be?  Is this an outcome you want? 
    \item What can NSF do?  What can the Resource Providers do?  
    \item How big a challenge is supporting constantly changing hardware platforms for you as a software provider? 
\end{itemize}
\subsubsection{Background}
A persistent challenge across CSSI tool sustainability is adapting the constant innovation in computing hardware. In recent years for example this has encompassed the transition to the universal use of GPUs for providing the bulk of the computing resources in HPC. 
New hardware clearly adds costs to supporting software as do updates in the broader software environment, vendor releases, and related factors. Strategies for sustaining and maintaining CSSI-supported software is the core issue addressed in this breakout.

\subsubsection{Current Status}
Summary themes from breakout:
\begin{itemize}
\item There is no clear path to get SW out through the RPs - find people you know at each one and ask is the status quo. 
\item Can NSF help by raising accountability and reporting?
\begin{itemize}
\item Like PAR for SW
\item OAC values SW… other divisions don’t!  Get to that “first class” Ben Brown proposed – grants are all “new”, so track record in successful SW doesn’t matter. 
\item If we want good SW, hold people accountable for those usability metrics. 
\end{itemize}
\item Link more clearly to papers for discoverability 
\item POSE/Sustainability track 
\begin{itemize}
\item In a grant to develop SW, 3 years long, you barely have a viable product before the end, and yet you are expected to push it, do outreach etc.
\item Those grant tracks are perceived as “terminal funding” 
\end{itemize}
\item Churn of HW/support is painful (postdoc lost 6 months with supercomputer transition at NCAR) 
\begin{itemize}
\item Though OS upgrades can be as painful as new HW. 
\end{itemize}
\item Even if we solve all this, networking, data, Inf. Sec. costs still will be huge problems. 
\begin{itemize}
\item One site could not release their software open source without clearing expensive institution-mandated information security reviews. \end{itemize} 
\item Where should the institution boundary be on solving these problems? Do we make it worse that every project crosses institutional boundaries?  
\item How do we balance demands on institutions with the need for diversity of institutions (i.e. avoid a system of “haves” and “have nots”). 
\end{itemize}

\subsubsection{Desired Outcomes / Future Directions}
\label{sec:breakout2_4-fut}
Create incentives for people to write better software, and reward the creation and maintenance of good software throughout government. 

\subsubsection{Recommendations}
\label{sec:breakout2_4-rec}

\begin{itemize}
\item Change the perception of POSE / CSSI-sustainability grants from “end of funding” to “next step in maintaining a healthy software base”.   
\item Create systems in reporting to value software like papers, track the requested metrics around adoption, etc., and hold PIs *accountable* for them (e.g. impacts on renewal or future SW funding). 
\item Spread the OAC culture of valuing software across NSF. 
\item Create a clearer path to transition CSSI-created software products throughout the NSF CI ecosystem (ACCESS, LCCF, etc.). 
\item Can we work with institutions to create an ecosystem to help support research software? (e.g. underlying expectations in release procedures, network infrastructure, etc.).  
\end{itemize}

\subsection{Community Building \& Measurable Broader Impacts }
\label{sec:breakout-community-impacts}
\label{sec:breakout2_5}
%
%
\paragraph{Objective} This breakout was focused on community building,and measures of the resulting broader impact with a careful description of current states, future impacts and recommendations to attain desired goals.
\footnote{Notes can be found here: \href{https://docs.google.com/document/d/1cAzFVpJQ6_L_EpJyAKREM7dUteg-zYzr7vPeSi1z9pI/edit?tab=t.0}{[meeting notes]}}

\subsubsection{Background}
Broader impacts extend from the advancement of scientific research and societal outcomes. Broader impacts achieved by building communities are a core tenet of the NSF CSSI, Cybertraining, and SCIPE programs. These three programs target different outcomes and, as such, build and support communities with different interests and goals. Although this offers several possibilities in extending Broader Impacts and societal outcomes, meaningfully measuring their outcomes continues to remain a challenge for these programs. 

\subsubsection{Current Status} 
The National Science Foundation (NSF) expects researchers' work to have broader impacts: the potential to benefit society and contribute to the achievement of specific, desired societal outcomes. Practices promoting broader impacts have come a long way. Perhaps we are approaching a set of minimum expectations. 

 The NSF is intentionally not prescriptive about the societal outcomes a project addresses, but provides examples of broader impacts across several categories. These include inclusion, STEM education, public participation, societal well-being, STEM workforce, partnerships, national security, economic competitiveness, infrastructure. As such, the communities in question can take different forms. Among others, the broader impacts in these communities can be achieved through developer communities, research support structures, curricular adoption, and informal training.  

 Practices to improve the broader impacts in a community are typical today. For example, transitioning ad hoc (or informal) training to the classroom and sharing curricular materials are increasingly considered standard practices. Today, they need to be considered with the same intentionality as science. It is not sufficient to simply mention activities. When considering broader impacts, researchers should consider the objectives, activities, and budgetary considerations.

Researchers are actively collecting data on extended research outcomes such as publications, software developed, students graduating, funding success, and courses developed. We note that while there are public tools, they may be  hard to use because they might expose student names, FERPA  issues.
 
Depending on their scope and funding, communities can take years to build. It difficult to gauge the success of community development over the lifetime of a shorter grant. Collecting metrics is somewhat easier in software development projects. Smaller programs are not appropriately funded to perform a true evaluation. Some schools have teams that assist with evaluations and collect metrics.  These facilities are, however, not available to all.

\subsubsection{Future Directions}
\label{sec:breakout2_5-fut}

The programs in this cohort are developing communities that focus on three critical albeit different aspects of advancing computationally enabled research. CSSI programs are actively developing communities of developers, and researchers who can adopt them in their workflows. The Cybertraining communities focus on the development and adoption of training materials in informal and formal settings. The SCIPE programs are focused on developing communities that can support researchers effectively use computing technologies.  

 The communities in this cohort  will have some overlapping needs, support structures, and ambitions. With this in mind, these communities can morph into others. They will further impact other communities downstream. As such, metrics have to consider the impact on secondary and tertiary communities. The community notes that these impacts may come into play after the duration of the original program. A prescriptive  mechanism that tracks the outcomes and objectives of a community could delineate its progress, but runs into the danger of becoming restrictive. 

 Reporting on the Broader Impacts is ingrained in most communities. While these three programs try to capture metrics such as demographics, attendance, publications from their communities, there is a need for more enhanced metrics. Participants in these communities engage for different reasons. While some communities are transient, others may exist for longer. One may consider how we can measure the sense of belonging in a community.  Perhaps a mechanism that supports a longitudinal study could capture this information. 

 There is an opportunity to elevate the discussion on what we are collecting. For example, there is a deeper need for mechanisms that help researchers develop, adopt and share new scientific practices. Unlike research outcomes, it is harder to capture this need in metrics.  While we are focused on advancing the use of metrics in a community, mechanisms to protect the participating researchers' privacy, and intellectual property need to be considered as well. Other programs may have developed unique ways to address this challenge. Here, specialized expertise for creating and evaluating approaches might prove helpful.  A group structured to meaningfully assist researchers engaging in these communities collectively could be helpful.  

\subsubsection{Recommendations}
\label{sec:breakout2_5-rec}

\paragraph{Refine and Expand Metrics}
Collaborate with social scientists to develop new metrics that capture both quantitative and qualitative aspects of community engagement. Consider what tools can analyze community sentiment, diversity, and long-term engagement trends. Allow communities to define what matters to them. 
\paragraph{Innovate Broader Impact Strategies}
Encourage original thinking in broader impact planning. Ensure appropriate budget allocations for long-term community engagement and sustainability initiatives. Explore the use of unstructured surveys to gain deeper insights. 
\paragraph{Strengthen Privacy and Legal Compliance}
Work with legal experts to ensure all tools and methods used in community-building are compliant with Federal, State and privacy laws. Consider adopting practices like partial name storage or anonymization to protect personally identifiable information while maintaining functionality.  Community leads must make a habit of requesting consent for all personal information collected and leveraging their institutional review boards (IRBs).
\paragraph{Invest in Expertise}
There is a need for a coordinated effort that brings expertise together in a cohesive manner. 
Mechanisms could include hiring or consulting with social scientists or other specialists in community engagement to guide the development of community-building initiatives. Provide training and resources to investigators to help them better understand and implement these strategies. 


\subsection{Sustainability and Continuing Training  }
\label{sec:breakout-sust-train}
\label{sec:breakout2_6}
\subsubsection{Background} 
The NSF Cybertraining program has been very successful.  The program has evolved from supporting training programs for users of CI systems to the introduction of the SCIPE program which focuses on the development of and training for CI professionals.  With that success comes the question of sustainability and scalability of the training programs and initiatives which have been supported.  
The 2023 Cybertraining PI workshop report included a recommendation to further define and recommend strategies for sustainability and scalability of program initiatives which prompted the development of this breakout session.

\subsubsection{Objective}
Define and recommend strategies for sustainability and the focus of the Cybertraining program for the future.  
   
\subsubsection{Current Status}
Sustainability and continuation of the NSF Cybertraining and SCIPE program is an important topic and concern of the community.  The program has been focused on curriciulum, training (short-term) and NSF directorate priorities.  There is an opportunity to expand from small scale projects to a national scope or domain-wide integration and regional collaboration.  There are limited industry partnerships and integration with the NSF TIP directorate which is seen as a missed opportunity for sustainability. The majority of the focus has been on college students (graduate).

\subsubsection{Suggested Questions} 
Where has Cybertraining been focused?  Where should it focus in the future?  Communities (i.e. college students, PIs, CI Professionals), topics, artifacts, curriculum
What is the future of Cybertraining? How do we scale, sustain and coordinate, gain efficiencies?  What is the role of industry (NAIRR, NSF TIP directorate)?  What components are needed for a successful center, institute?

\subsubsection{Desired Outcomes / Future Directions}
\label{sec:breakout2_6-fut}
A sustainable resource and personnel ecosystem for training and education of the cyberscientists, RSEs and associated resources without {\it ad hoc} and one-time grant funding. Training pathways, career definitions and business models for both are well defined and invested in.

\subsubsection{Recommendations}
\label{sec:breakout2_6-rec}
\paragraph{Structured Organization and Community Building}
Develop an 'Alliance' model which fosters broader partnerships, bringing projects together under collaborative umbrellas, similar to Research Coordination Networks (RCNs), BigData Hubs, AI institutes. Alliances and hubs, which may be discipline or regionally focused, will share resources and best practices via a central repository to avoid duplication of efforts.  Regional collaboration can foster the sharing of cybertraining-proficient people for scalability. Create a digital library of repository materials and share formal curricula through this federated repository. 
\paragraph{Community-driven and Innovative Curriculum}
The cyberinfrastructure community must adapt to rapidly changing technology by ensuring that materials and delivery methods remain up to date and relevant.  Community-driven curriculum and topics are important to maintain our competitive advantage.  Programs similar to the Software Carpentries program should be considered to teach cyberinfrastructure to scientists, specifically cloud based infrastructure as a service.  Programs should expand their target audiences to include K-1, community colleges and current CI professional from academia and industry.  Programs should be considered which teach professional skills in addition to technical skills, include effective communication, project management and cybersecurity/privacy awareness. 
\paragraph{Inter-directorate and Industry Collaboration}
It was suggested that the development of two-way relationships with other NSF directorates, such as EDU and domain directorates, would be useful, specifically incorporating EDU pedagogy research into Cybertraining awards.  A strong recommendation from this year's participants and last year's participants is to partner with the NSF TIP directorate and industry partners to develop training which is valuable to industry for staff upskilling providing an opportunity for subsidies and sustainability in addition to other opportunities.

\section{Acknowledgements}
This workshop was made possible with NSF Awards 435580 and 2434556. We are grateful to NSF for helping to arrange the collaborative CSSI PI and CyberTraining/SCIPE PI meetings. We also thank the following NSF Program Officers who were present: Sonam Ahluwalia, Amy Apon, Sharmistha Bagchi-Sen, Linkan Bian, Varun Chandola, Sharon Geva, Sheikh Ghafoor, Tom Gulbransen, Marlon Pierce, Plato Smith, Ashok Srinivasan, Wen-wen Tung, Rediet Woldeselassie, Sam Xin, May Yuan. We also want to acknowledge the efforts of the following individuals, whose contributions often went above and beyond the call of duty, and this meeting would simply not have happened, including: Susan Rathbun (SDSC); Jordan Wilkinson and Meghan Rodriguez (Tufts); and Hantao Cui (NCSU) for developing and hosting the event website.
\label{sec:acknowledgements}

\bibliographystyle{unsrt}  
\bibliography{cytr-cssi-24}

\pagebreak
\appendix 

\newpage
\section{Selected Slides \& Presentations}

\subsection{NSF Funding Opportunities}
\includepdf[pages=-]{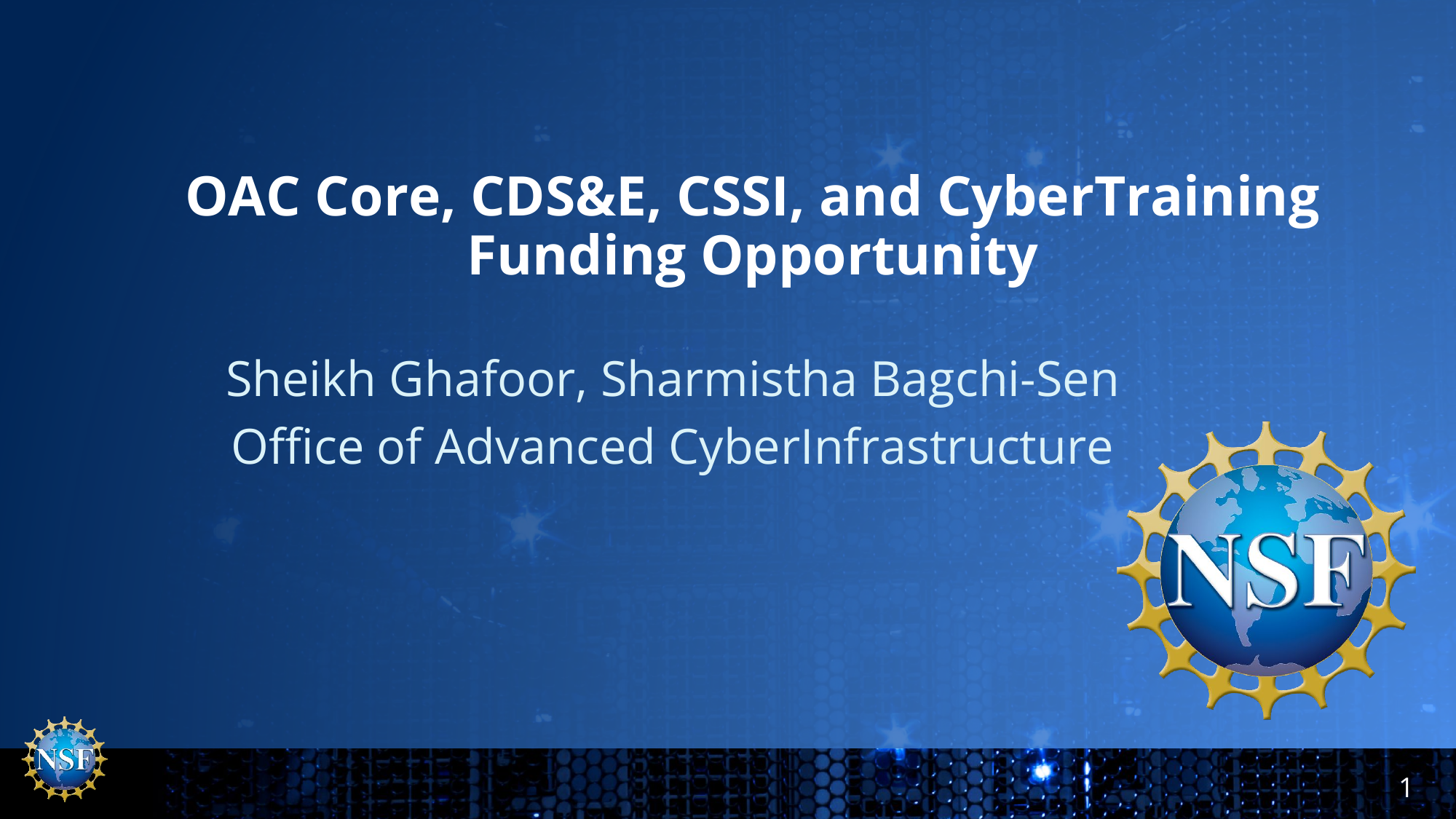}

\subsection{NSF ACCESS Resources and Opportunities}
\includepdf[pages=-]{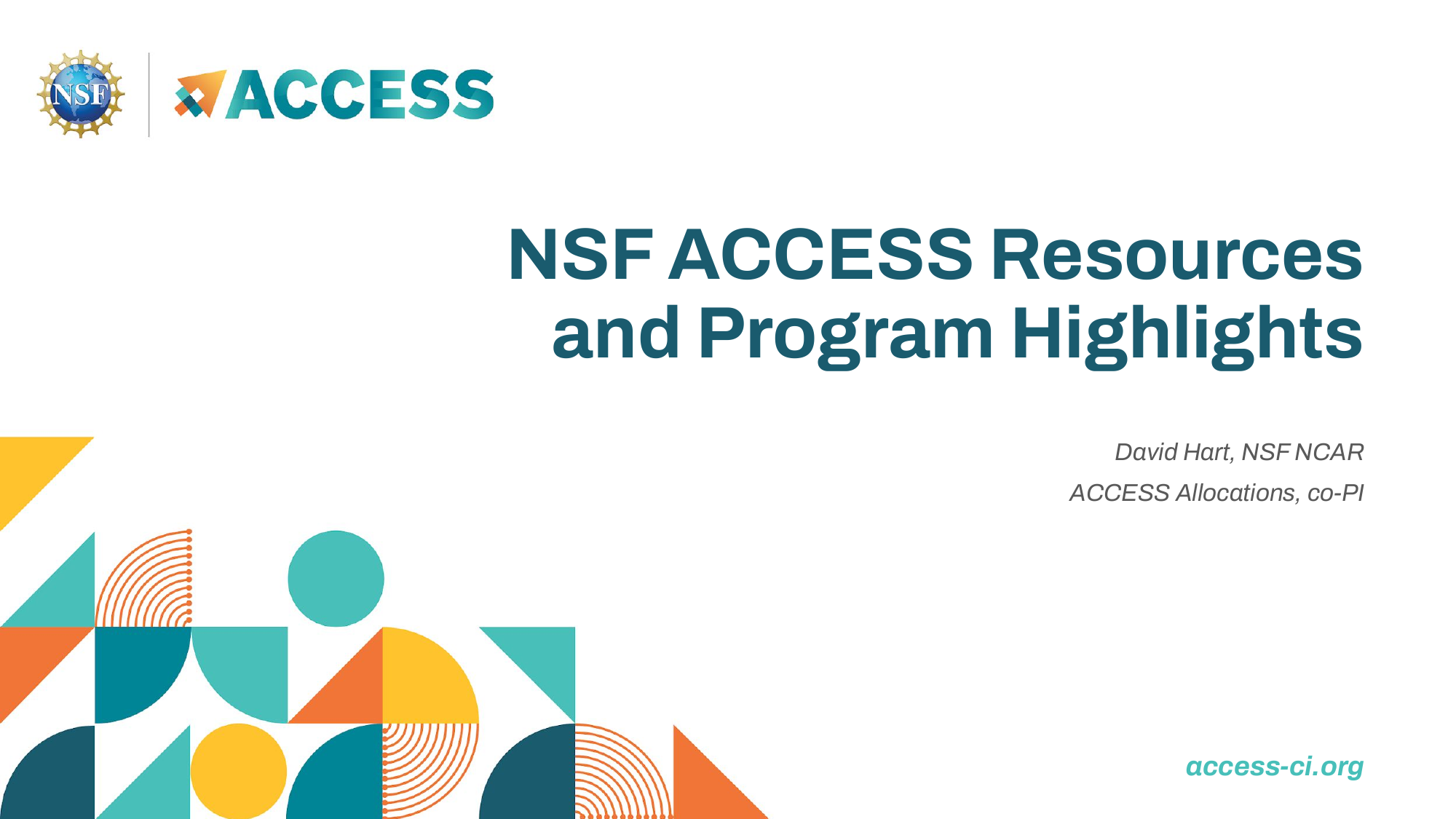}

\subsection{Metrics for Cybertraining/SCIPE Programs and Outreach}
\includepdf[pages=-]{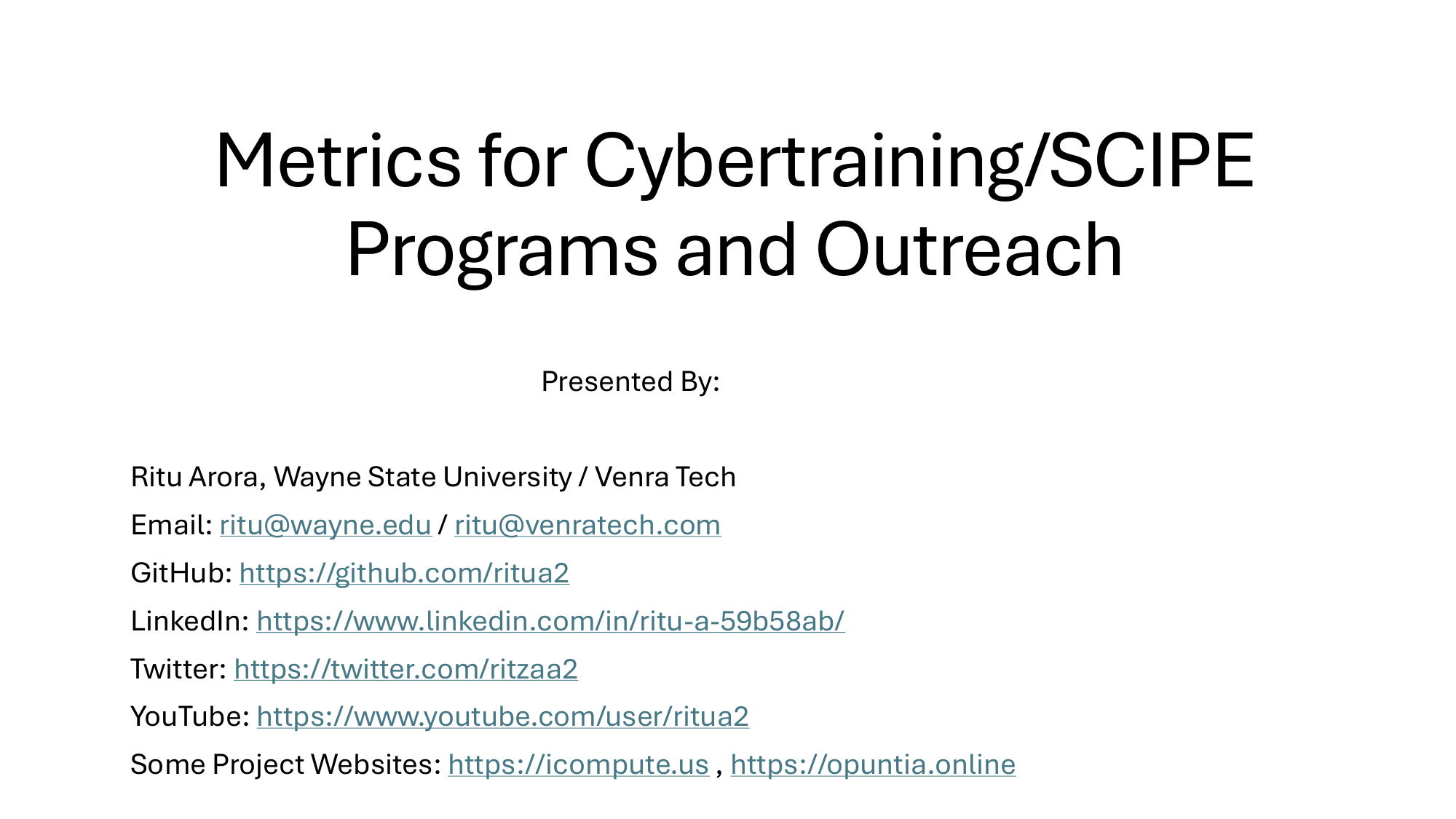}

\subsection{Promoting Better Scientific Software}
\includepdf[pages=-]{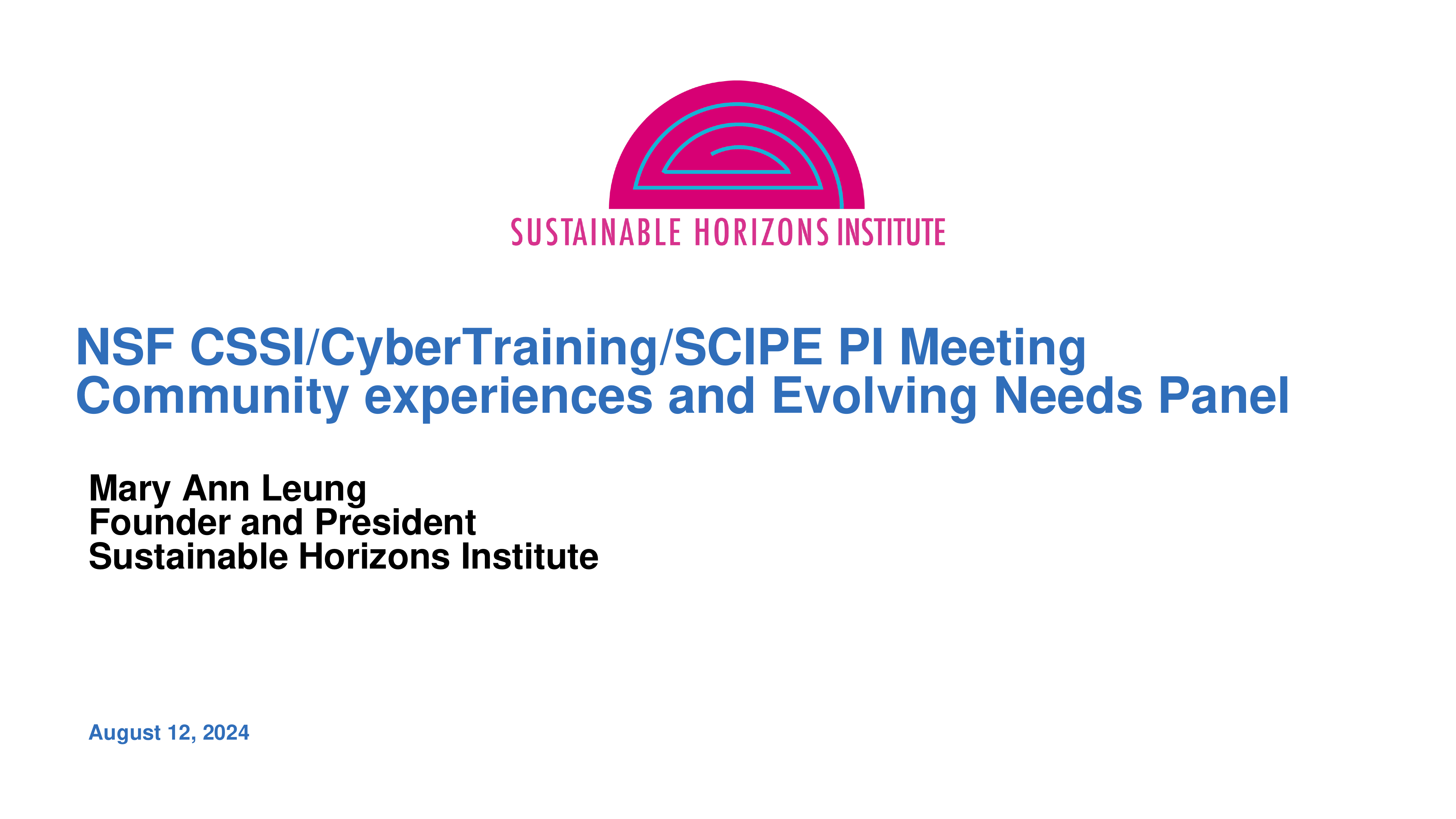}

\subsection{Democratizing Science through Cyberinfrastructure}
\includepdf[pages=-]{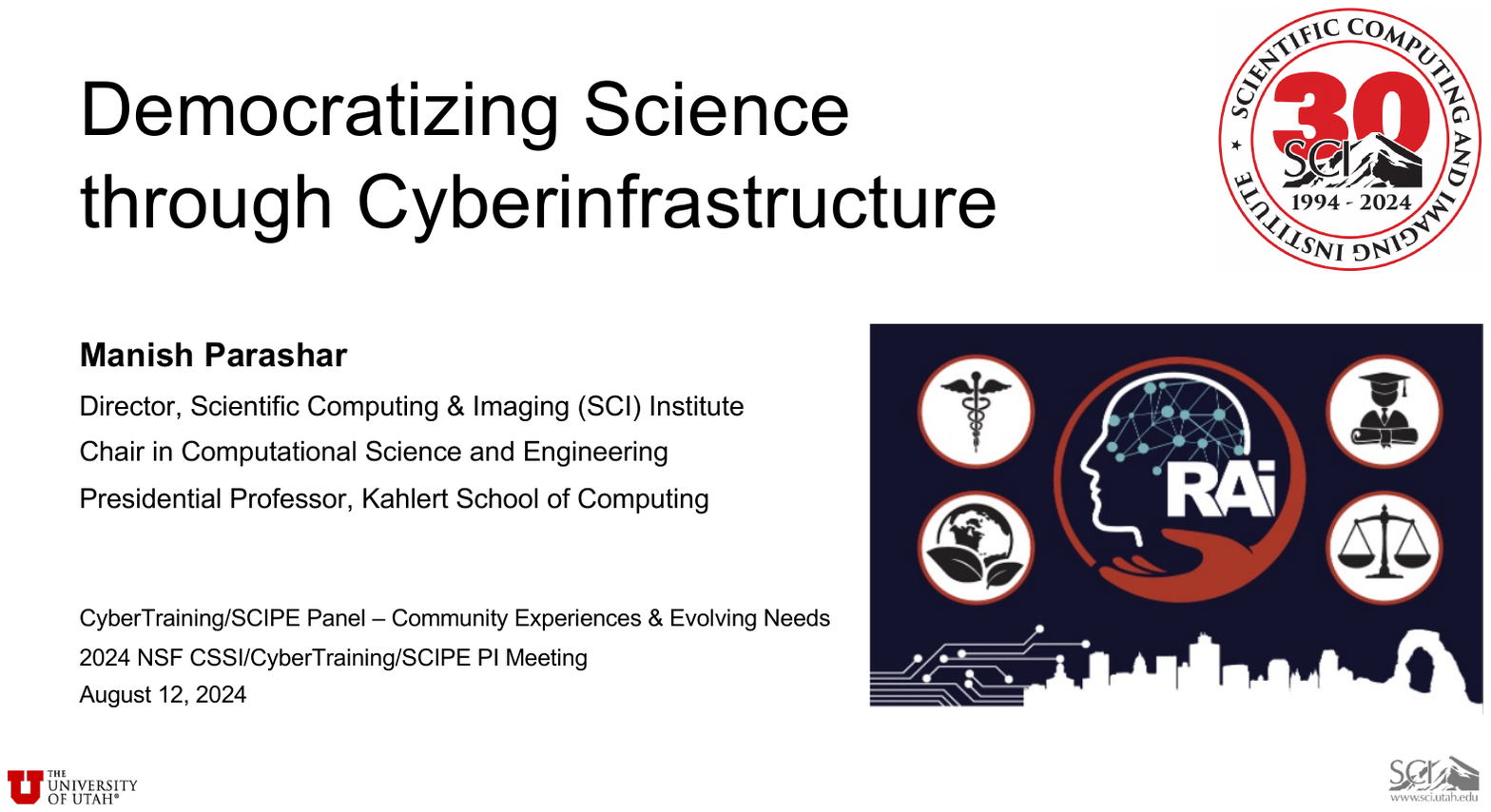}

\newpage
\section{Participant List, Poster List \& DOIs}

Overall, there were 286 participants, representing 292 awards funded by CSSI, CyberTraining, OAC Core, CIP, SCIPE CDS \& E and related programs. 
Among the participants, the following NSF Program Directors were present: Sonam Ahluwalia, Amy Apon, Sharmistha Bagchi-Sen, Linkan Bian, Varun Chandola, Sharon Geva, Sheikh Ghafoor, Tom Gulbransen, Marlon Pierce, Plato Smith, Ashok Srinivasan, Wen-wen Tung, Rediet Woldeselassie, Sam Xin, May Yuan.


\begin{sidewaystable}
\begin{center}
\resizebox{\textheight}{!}{
\begin{tabular}{ |l |l |l |l |l  }
{\bf PARTICIPANT LAST NAME} & {\bf FIRST NAME} & {\bf AWARD \#} & {\bf AWARD TITLE} & {\bf Company} \\
\hline

Agar & Joshua & 2320600 & MRI: Track 2 Development of a Platform for Accessible Data-Intensive Science and Engineering & Drexel University \\
Akimov & Alexey & 1924256 & CyberTraining: Pilot: Modeling Excited State Dynamics in Solar Energy Materials & University at Buffalo, SUNY \\
Akimov & Alexey & 1931366 & Elements: Libra: The Modular Software for Nonadiabatic and Quantum Dynamics & University at Buffalo, SUNY \\
Akli & Linda & 2400201 & CIU: Implementation: Medium: Computational and Data Science Curriculum Exchange Faculty Community of Practice & Texas Advanced Computing Center \\
Aksamija & Zlatan & 2302879 & CDSE: Coupled Electro-Thermal Transport in Two-Dimensional Materials and Heterostructures & University of Utah \\
Amrhein & Dan & 2311382 & Collaborative Research: Frameworks: A community platform for accelerating observationally-constrained regional oceanographic modeling & NSF NCAR \\
Andreussi & Oliviero & 2321102 & Collaborative Research: CyberTraining: Implementation: Medium: Training Users, Developers, and Instructors at the Chemistry/Physics/Materials Science Interface & Boise State University \\
\end{tabular}
}
\end{center}
\end{sidewaystable}

\pagebreak
\begin{sidewaystable}
\begin{center}
\resizebox{\textheight}{!}{
\begin{tabular}{ |l |l |l |l |l  }
{\bf PARTICIPANT LAST NAME} & {\bf FIRST NAME} & {\bf AWARD \#} & {\bf AWARD TITLE} & {\bf Company} \\
\hline
Eisma & Jessica & 2230054 & CyberTraining: Pilot: Justice in Data: An intensive, mentored online bootcamp developing FAIR data competencies in undergraduate researchers in the water and energy sectors & University of Texas at Arlington \\
Esfarjani & Keivan & 2103989 & ELEMENTS: Anharmonic formalism and codes to calculate thermal transport and phase change from first-principles calculations & University of Virginia \\
Evans & John & 2104106 & Collaborative Research: Elements: EXHUME: Extraction for High-Order Unfitted Finite Element Methods & University of Colorado Boulder \\
Figueiredo & Renato & 2311123 & Collaborative Research: Elements: FaaSr: Enabling Cloud-native Event-driven Function-as-a-Service Computing Workflows in R & University of Florida \\
Figueiredo & Renato & 2311124 & Collaborative Research: Elements: FaaSr: Enabling Cloud-native Event-driven Function-as-a-Service Computing Workflows in R & University of Florida \\
Foster & Ian & 2209892 & Frameworks: Garden: A FAIR Framework for Publishing and Applying AI Models for Translational Research in Science, Engineering, Education, and Industry & University of Chicago and Argonne National Laboratory \\
Ganapati & Sukumar & 1924154 & CyberTraining: Implementation: Medium: Advanced Cyber Infrastructure Training in Policy Informatics & Florida International University \\
\end{tabular}
}
\end{center}
\end{sidewaystable}

 \pagebreak
\begin{sidewaystable}
\begin{center}
\resizebox{\textheight}{!}{
\begin{tabular}{ |l |l |l |l |l  }
{\bf PARTICIPANT LAST NAME} & {\bf FIRST NAME} & {\bf AWARD \#} & {\bf AWARD TITLE} & {\bf Company} \\
\hline
Machi & Dustin & 1916805 & CINES: A Scalable Cyberinfrastructure for Sustained Innovation in Network Engineering and Science & University of Virginia \\
Maffeo & Christopher & 2311550 & Elements: Enabling multi-resolution simulations at the interface of biology and nanotechnology with ARBD & University of Illinois at Urbana-Champaign \\
Mandal & Subhasish & 2311558 & Collaborative Research: Elements: Building an open source DFT+eDMFT database for quantum materials & West Virginia University \\
Mandal & Anirban & 2320373 & Collaborative Research: CyberTraining: Implementation: Medium: CyberInfrastructure Training and Education for Synchrotron X-Ray Science (X-CITE) & UNC Chapel Hill \\
Manning & T. Andrew & 2103680 & Frameworks: MUSES, Modular Unified Solver of the Equation of State & National Center for Supercomputing Applications, University of Illinois Urbana-Champaign \\
Marru & Suresh & 2209872 & Collaborative Research: Frameworks: Cybershuttle: An end-to-end Cyberinfrastructure Continuum to accelerate Discovery in Science and Engineering & Georgia Tech \\
Martiniani & Stefano & 2311632 & GOALI: Frameworks: At-Scale Heterogeneous Data based Adaptive Development Platform for Machine-Learning Models for Material and Chemical Discovery & New York University \\
\end{tabular}
}
\end{center}
\end{sidewaystable}

 \pagebreak
\begin{sidewaystable}
\begin{center}
\resizebox{\textheight}{!}{
\begin{tabular}{ |l |l |l |l |l  }
{\bf PARTICIPANT LAST NAME} & {\bf FIRST NAME} & {\bf AWARD \#} & {\bf AWARD TITLE} & {\bf Company} \\
\hline
Taha & Ahmad & 2230087 & Collaborative Research: CyberTraining: Implementation: Medium: Cross-Disciplinary Training for Joint Cyber-Physical Systems and IoT Security & Vanderbilt University \\
Takabi & Daniel & 2413654 & CyberTraining: Implementation: Small: Building Future Research Workforce in Trustworthy Artificial Intelligence (AI) & Old Dominion University \\
Taylor & Greg & 2103707 & Collaborative Research: Elements: A CPU/GPU Pipeline Framework for High Throughput Data Acquisition and Analysis & The University of New Mexico \\
Thiruvathukal & George & 2104319 & CDSE: Collaborative: Cyber Infrastructure to Enable Computer Vision Applications at the Edge Using Automated Contextual Analysis & Loyola University Chicago \\
Thomas & Mary & 2017767 & CyberTraining: Developing a Best Practices Training Program in Cyberinfrastructure-Enabled Machine Learning Research & University of California at San Diego \\
Thomas & Mary & 2230127 & CIP: Training and Developing a Research Computing and Data CI Professionals (RCD-CIP) Community & University of California at San Diego \\
Tian & Yuan & 2411153 & Collaborative Research: Frameworks: MobilityNet: A Trustworthy CI Emulation Tool for Cross-Domain Mobility Data Generation and Sharing towards Multidisciplinary Innovations & University of California Los Angeles \\
\end{tabular}
}
\end{center}
\end{sidewaystable}

\pagebreak
\newpage 



\pagebreak
\begin{sidewaystable}
\begin{center}
\resizebox{\textheight}{!}{
\begin{tabular}{ |l |l |l |l   }
{\bf AUTHORS: A to Z} & {\bf AWARD\#} & {\bf POSTER TITLE} & {\bf figshare DOI} \\
\hline

Name & 12345 & TItle & figshareDOI  \\
 Agar, Joshua & 2209135 & Cyberinfrastructure for Scientific Data Preservation and Image Similarity Search & 10.6084/m9.figshare.26798377 \\
Akimov, Alexey  & 1924256 & CyberTraining: Pilot: Modeling Excited State Dynamics in Solar Energy Materials & 10.6084/m9.figshare.26532907.v1 \\
Akimov, Alexey & 1931366 & Elements: Libra: The Modular Software for Nonadiabatic and Quantum Dynamics & 10.6084/m9.figshare.26532907.v1 \\
Akli, Linda & 2400201 & Computational and Data Science Curriculum Exchange (CDS-Exchange) & 10.6084/m9.figshare.26517091.v1 \\
Aksamija, Zlatan & 2302879 & Thermal and Electronic Transport Dynamics in 2D Materials & 10.6084/m9.figshare.26800810.v1 \\
Amrhein, Dan & 2311382 & CROCODILE: Facilitating rapid prototyping and analysis of regional ocean physics and carbon cycle modeling with data assimilation & 10.6084/m9.figshare.26820904 \\
\end{tabular}
}
\end{center}
\end{sidewaystable}

\pagebreak
\begin{sidewaystable}
\begin{center}
\resizebox{\textheight}{!}{
\begin{tabular}{ |l |l |l |l   }
{\bf AUTHORS: D to K} & {\bf AWARD\#} & {\bf POSTER TITLE} & {\bf figshare DOI} \\
\hline

D'Orazio, Vito & 2311142 & Frameworks: Infrastructure For Political And Social Event Data using Machine Learning & 10.6084/m9.figshare.26720437 \\
Dabaghmeshin, Mahsa &  2229652 & CIberCATSS: A Comprehensive, Applied and Tangible CyberInfrastructure
Summer School in Southeastern Wisconsin & 10.6084/m9.figshare.26513080 \\
Davidson, Drew & 2311819 & Prototyping a Quality Control / Quality Assurance System for the StraboSpot Geologic Information System & 10.6084/m9.figshare.26767156.v1 \\
DePrince, Eugene & 2103705 & Unitary Coupled Cluster in the Chronus Quantum Package & 10.6084/m9.figshare.26768125.v1 \\
Denolle, Marine & 2103701 & SCOPED Update: a Cloud and HPC software platform for computational seismology &  10.6084/m9.figshare.26513095.v2 \\
Deppe, Nils & 2209655 & SpECTRE: A task-based code for multiphysics problems in astrophysics at exascale & 10.6084/m9.figshare.26798689 \\
Di, Sheng & 2311875 & FZ: A fine-tunable cyberinfrastructure framework to streamline specialized lossy compression development & 10.6084/m9.figshare.26359654.v1 \\
\end{tabular}
}
\end{center}
\end{sidewaystable}

\pagebreak
\begin{sidewaystable}
\begin{center}
\resizebox{\textheight}{!}{
\begin{tabular}{ |l |l |l |l   }
{\bf AUTHORS: L to O} & {\bf AWARD\#} & {\bf POSTER TITLE} & {\bf figshare DOI} \\
\hline

Lange, David & 2311471 & Clad: Enabling Differentiable Programming in Science & 10.6084/m9.figshare.26780410 \\
Lejeune, Emma & 2310771 & Curating and Disseminating Solid Mechanics Based Benchmark Datasets & 10.6084/m9.figshare.26540602 \\
Lele, Sanjiva & 2103509 & Compack3D: Accelerating high-order compact scheme simulations & 10.6084/m9.figshare.28138154 \\
Lemson, Gerard & 2311791 & Open SciServer: A Sustainable Data-Driven Science Platform & 10.6084/m9.figshare.26568826.v1 \\
Lewis, Thu-Mai & 2209628, 2209629, 2209630 & TRAnsparency CErtified (TRACE): Trusting Computational Research Without Repeating It & 10.6084/m9.figshare.26530711 \\
Li, Xiaosong & 2103717 & Sustainable Open-Source Quantum Dynamics and
Spectroscopy Software & 10.6084/m9.figshare.26768125.v1 \\
Li, Dong & A210125002 & Accelerating LLM Inference Using Memoization in SciMem & 10.6084/m9.figshare.26384722 \\
\end{tabular}
}
\end{center}
\end{sidewaystable}

\pagebreak
\begin{sidewaystable}
\begin{center}
\resizebox{\textheight}{!}{
\begin{tabular}{ |l |l |l |l   }
{\bf AUTHORS: P to S} & {\bf AWARD\#} & {\bf POSTER TITLE} & {\bf figshare DOI} \\
\hline

Pallickara, Shrideep & 1931363, 1931324, 1931335,,1931283  & SUSTAIN: Catalyzing Urban Sustainability At Scale & MISSING \\
Panda, Dhabaleswar  & 2311830 & CSSI Frameworks: Performance engineering scientific applications with MVAPICH and TAU using emerging communication primitives & 10.6084/m9.figshare.26841616 \\
Panda, Dhabaleswar  & 2007991 & OAC Core: Next-generation communication and I/O middleware for HPC and deep learning with smart NICs & 10.6084/m9.figshare.26841616 \\
Pantano-Rubino, Carlos & 2320943 & Fostering Computational Excellence (FOCEX): Addressing the Disconnect between Advanced CyberInfrastructure and Educational Preparedness & 10.6084/m9.figshare.26543671 \\
Park, Seung-Jong & 2403399 & Harvesting Idle Resources Safely and Timely for Large-scale AI Applications in High Performance Computing Systems & 10.6084/m9.figshare.26499400.v1 \\
Patel, Vivak & 2309445 & A cyberlaboratory for randomized linear algebra & 10.6084/m9.figshare.26591698 \\
Patra, Abani & 2004826 & Ghub : Building a glaciology gateway to unify a community & MISSING \\
\end{tabular}
}
\end{center}
\end{sidewaystable}

\pagebreak
\begin{sidewaystable}
\begin{center}
\resizebox{\textheight}{!}{
\begin{tabular}{ |l |l |l |l   }
{\bf AUTHORS: T to Z} & {\bf AWARD\#} & {\bf POSTER TITLE} & {\bf figshare DOI} \\
\hline
Taha, Ahmad & 2230087 & Stealthy False Data Injection Attacks on Drinking Water and Power Systems & 10.6084/m9.figshare.26662243 \\
Tajkhorshid, Emad & 2139536 & Boosting the Performance of NAMD on GPU-enabled Platforms & MISSING \\
Takabi, Daniel & 2413654 & Building Future Research Workforce in Trustworthy Artificial Intelligence (AI) & 10.6084/m9.figshare.26799499 \\
Tang, Yufei & 2320972 & Secure, Resilient Cyber-Physical Energy System Workforce Pathways via Data-Centric, Hardware-in-the-Loop Training & 10.6084/m9.figshare.28138145 \\
Taylor, Greg & 2103707 & Using Bifrost for High Speed GPU Imaging Pipelines & 10.6084/m9.figshare.26536159 \\
Thiruvathukal, George & 2104319 & Cyber Infrastructure to Enable Computer Vision Applications at the Edge Using Automated Contextual Analysis & 10.6084/m9.figshare.26530858 \\
Thiruvathukal, George & 2107020 &  Advancing Low-Power Computer Vision at the Edge & 10.6084/m9.figshare.26530858 \\
\end{tabular}
}
\end{center}
\end{sidewaystable}

\pagebreak

\end{document}